\documentclass[pra,twocolumn,showkeys,showpacs,nofootinbib,superscriptaddress,10pt
]{revtex4-1}
\usepackage{graphicx}  
\usepackage{grffile}
\usepackage{amsthm}
\usepackage{epsfig}
\usepackage{amsmath,amssymb,amsfonts,mathtools,bbm,MnSymbol,mathrsfs,xfrac}   
\usepackage{longtable} 
\usepackage{tabularx}
\usepackage{colortbl}
\usepackage[table]{xcolor}
\usepackage[breaklinks=true,colorlinks=true,linkcolor=blue,urlcolor=blue,citecolor=blue]{hyperref}
\usepackage{comment}
\usepackage{color}
\usepackage[makeroom]{cancel}
\usepackage{footmisc}

\usepackage{soul}

\newcommand{\CN}{\mathcal{N}}

\renewcommand{\[}{\begin{equation}}
\renewcommand{\]}{\end{equation}}

\newcommand{\ket}[1]{|#1\rangle}
\newcommand{\bra}[1]{\langle#1|}
\newcommand{\braket}[2]{\langle#1|#2\rangle}
\newcommand{\pro}[2]{|#1\rangle\langle#2|}
\newcommand{\mean}[1]{\langle#1\rangle}
\newcommand{\abs}[1]{|#1|}

\newcommand{\tr}{\mathrm{tr}}

\newcommand{\R}{{\hat{\rho}}}

\newcommand{\C}{{\mathcal{C}}}
\renewcommand{\P}{\hat{P}}
\newcommand{\bi}{{\boldsymbol{i}}}

\newcommand{\HS}{\mathcal{H}}

\newcommand{\be}{\boldsymbol{\epsilon}}
\newcommand{\bx}{{\boldsymbol{x}}}
\newcommand{\bp}{{\boldsymbol{p}}}
\newcommand{\bn}{{\boldsymbol{n}}}
\newcommand{\bE}{{\boldsymbol{E}}}

\newcommand\restr[2]{{
  \left.\kern-\nulldelimiterspace 
  #1 
  \vphantom{\big|} 
  \right|_{#2} 
  }}

\newcommand{\ds}[1]{\textcolor{red}{[{\bf DS}: #1]}}
\definecolor{mygray}{gray}{0.6}

\theoremstyle{definition}
\newtheorem{definition}{Definition}
\newtheorem{theorem}{Theorem}

\begin{document}

\title{Classical dynamical coarse-grained entropy\\
and comparison with the quantum version
}

\author{Dominik \v{S}afr\'{a}nek}
\email{dsafrane@ucsc.edu}
\affiliation{SCIPP and Department of Physics, University of California, Santa Cruz, CA 95064, USA}
\author{Anthony Aguirre}
\affiliation{SCIPP and Department of Physics, University of California, Santa Cruz, CA 95064, USA}
\author{J. M. Deutsch}
\affiliation{Department of Physics, University of California, Santa Cruz, CA 95064, USA}

\date{\today}

\begin{abstract}
We develop the framework of classical Observational entropy, which is a mathematically rigorous and precise framework for non-equilibrium thermodynamics, explicitly defined in terms of a set of observables. Observational entropy can be seen as a generalization of Boltzmann entropy to systems with indeterminate initial conditions, and describes the knowledge achievable about the system by a macroscopic observer with limited measurement capabilities; it becomes Gibbs entropy in the limit of perfectly fine-grained measurements. This quantity, while previously mentioned in the literature, has been investigated in detail only in the quantum case~\cite{safranek2019letter,safranek2019long,strasberg2019entropy,strasberg2020heat}.
We describe this framework reasonably pedagogically, then show that in this framework, certain choices of coarse-graining lead to an entropy that is well-defined out of equilibrium, additive on independent systems, and that grows towards thermodynamic entropy as the system reaches equilibrium, even for systems that are genuinely isolated. Choosing certain macroscopic regions, this dynamical thermodynamic entropy measures how close these regions are to thermal equilibrium. We also show that in the given formalism, the correspondence between classical entropy (defined on classical phase space) and quantum entropy (defined on Hilbert space) becomes surprisingly direct and transparent, while manifesting differences stemming from non-commutativity of coarse-grainings and from non-existence of a direct classical analogue of quantum energy eigenstates.
\end{abstract}

\maketitle

\section{Motivation}

From the introductory pedagogical level to contemporary fundamental research, and in both classical and quantum contexts, the concept of entropy has caused a great deal of confusion. Not only are there many definitions of different type of entropies appropriate to different contexts~\cite{jaynes1965gibbs,wehrl1978general,swendsen2011physicists,balibrea2016clausius,goldstein2017nonequilibrium,goldstein2019gibbs}, but, we would contend, there are two distinct fundamental {\em notions} of to what entropy is meant to refer.  

On one hand, entropy per the definitions of Shannon, Gibbs, or von Neumann, is an information-theoretic quantity associated with the probabilities attributed to states of a system.  This entropy is preserved in a closed system undergoing evolution via the classical Liouville equation or a unitary quantum operator, reflecting the preservation of information in such systems, but the entropy may change (and will generally rise) if interactions with an external system are allowed.

On the other hand, entropy can measure how ``generic"  some state-of-affairs is, as described at a coarse-grained or macroscopic level.  Thermodynamic entropy, and that defined by Boltzmann as the number of microstates associated with a given macrostate, share this character.  This sort of entropy can (and tends to) increase in a closed system, satisfying some version of the Second Law of thermodynamics.

These notions are often conflated because they tend to coincide in equilibrium systems, all converging to the logarithm of the number of states compatible with some set of fixed constraints -- whether those states are cells in classical state space, energy levels in quantum theory, or defined otherwise.  Yet if we wish to describe systems {\em out} of equilibrium, in which entropy can evolve, or to evaluate entropy for small numbers of particles, more conceptual and mathematical precision is necessary.

In this paper we argue that there is a natural and rigorous definition of entropy at the classical level that:
\begin{enumerate}
\item Is well-defined in any classical system with a fixed phase-space and probability measure over that phase-space; in particular it is well-defined out of equilibrium and for small numbers of particles.
\item Constitutes a generalization of, and an interpolation between, classical Gibbs and Boltzmann entropies.
\item Is defined in terms of a {\em coarse-graining} that corresponds to a partitioning of phase-space.
\item Evolves continuously in time, and generically toward larger values, corresponding to a Second Law.
\item Can be cleanly transcribed into the quantum context -- and in fact is the classical version of the quantum ``Observational entropy" introduced by~\cite{safranek2019letter,safranek2019long}.
\end{enumerate}
and in addition,
\begin{enumerate}
  \setcounter{enumi}{5}
\item Can be used to describe the dynamics of classical systems. Specifically, it can be used to define ``dynamical thermodynamic entropy'' that changes with the evolution of a system.
\end{enumerate}

The idea of coarse-graining has a very long history, going back to Boltzmann~\cite{boltzmann2015relationship}, and coarse-grained entropies or coarse-grained free energies have emerged in numerous applications, such as fluid dynamics and Navier-Stokes equations~\cite{espanol1997coarse,gao2017analytical} (clearly present in the \emph{continuum hypothesis}~\cite{batchelor1967introduction}), statistical mechanics of fields and renormalization group~\cite{fisher1998renormalization,kardar2007statisticalparticles,kardar2007statisticalfields,ma2018modern}, in chemical engineering in computing the entropy of mixing~\cite{smith1950introduction,guggenheim1956statistical,callen1998thermodynamics}, and in field theory in the guise of renormalization (leading to the 1982 Nobel Prize in Physics for work on critical phenomena using the renormalization group~\cite{wilson1971renormalization}).
And there are definitions of entropy using some type of rigorously-defined coarse-graining, such as entropy of partition~\cite{jost2006dynamical}, Kolmogorov-Sinai entropy~\cite{farmer1982information,latora1999kolmogorov,frigg2004sense,jost2006dynamical}, or topological entropy~\cite{farmer1982information,jost2006dynamical}. (These mostly apply to dynamical systems.)

Nonetheless, in many works coarse-graining is treated in a rather ad-hoc and non-rigorous manner, with many subtleties swept under the rug. The coarse-grained entropy that we are going to argue for here has also appeared previously in literature. The quantum version of this entropy predates the classical; it was first introduced by von Neumann~\cite{von2010proof,von1955mathematical} as a resolution to the fact that the [von Neumann] entropy does not increase in isolated systems, then briefly mentioned by Wehrl~\cite{wehrl1978general} as ``coarse-grained entropy.'' It was further developed and generalized to include multiple coarse-grainings by the present authors in~\cite{safranek2019letter,safranek2019long}, and  termed ``Observational entropy,'' because it can be interpreted as the amount of knowledge an observer has about the system, when they have access only to a set of coarse-grained measurements (and each measurement corresponds to an observable). Quantum Observational entropy has been found to dynamically describe thermalization of isolated quantum systems~\cite{safranek2019long,lent2019quantum}, has been discussed in relationship with other types of entropies~\cite{goldstein2019gibbs}, has been found to increase under Markovian stochastic maps~\cite{gemmer2014entropy}, has been argued for as a natural candidate for entropy production~\cite{strasberg2019entropy,strasberg2020heat} because its definition does not need an explicit temperature dependence, found to follow an integral fluctuation theorem and providing an alternative perspective on the conventional definition of mechanical work in quantum systems~\cite{strasberg2020heat}, and has been shown to illustratively model isolated systems out of equilibrium by studying its extreme fluctuations~\cite{faiez2019extreme}.

Classical Observational entropy has appeared previously~\cite{wehrl1978general,latora1999kolmogorov,nauenberg2004evolution,kozlov2007fine,piftankin2008gibbs,vzupanovic2018relation}, but generally just as a mention in passing,\footnote{Eq.~(1.26) in~\cite{wehrl1978general}, Eq.~(1) in~\cite{latora1999kolmogorov} (up to a minus sign and an additive constant), Eq.~(A.1) in~\cite{nauenberg2004evolution}, Eq.~(1.1) in~\cite{kozlov2007fine}, Def.~4.1 in~\cite{piftankin2008gibbs} and Eq.~(11) in~\cite{vzupanovic2018relation}.}  but never with a clear or comprehensive discussion of it, nor with any compelling treatment of applications to thermodynamics.

Thus the aim of this paper is to motivate and define a rigorous mathematical framework of classical Observational entropy, which we believe gives an elegant unification of many coarse-graining techniques, then derive its various properties, interpret it both from an information-theoretic perspective and from a physical perspective (by connecting it to thermodynamic entropy), and finally spell out its relations to the quantum version. In other words, we will both argue that our definition of coarse-grained entropy satisfies the desiderata in points $1$--$6$ above, and also provide the reader intuition about behavior of this entropy, by making the connection between the classical and quantum versions clear.

\section{Boltzmann entropy}

Although there are variations, the type of ``state counting" entropy defined by Boltzmann~\cite{boltzmann2015relationship,boltzmann2003further,brush2004history} generally attributes a number  $V$ of fundamental microstates to a given macrostate -- defined in some terms -- attributing an entropy proportional to $\ln V$ to the macrostate. To be more precise, we make use of the following definitions.

The state space of a system can be partitioned into non-overlapping subspaces that sum up to the full state space.  This partition is called a ``coarse-graining" and denoted by $\C$. An element of this partition is a called a macrostate.

For a microstate $m$ in a macrostate, we attach Boltzmann entropy of $S_B(m)=\ln(V)$ to this microstate (as well as to the macrostate of which it is an element) where $V$ is the number of microstates contained in the macrostate.

\begin{figure}[t]
\begin{center}
\includegraphics[width=1\hsize]{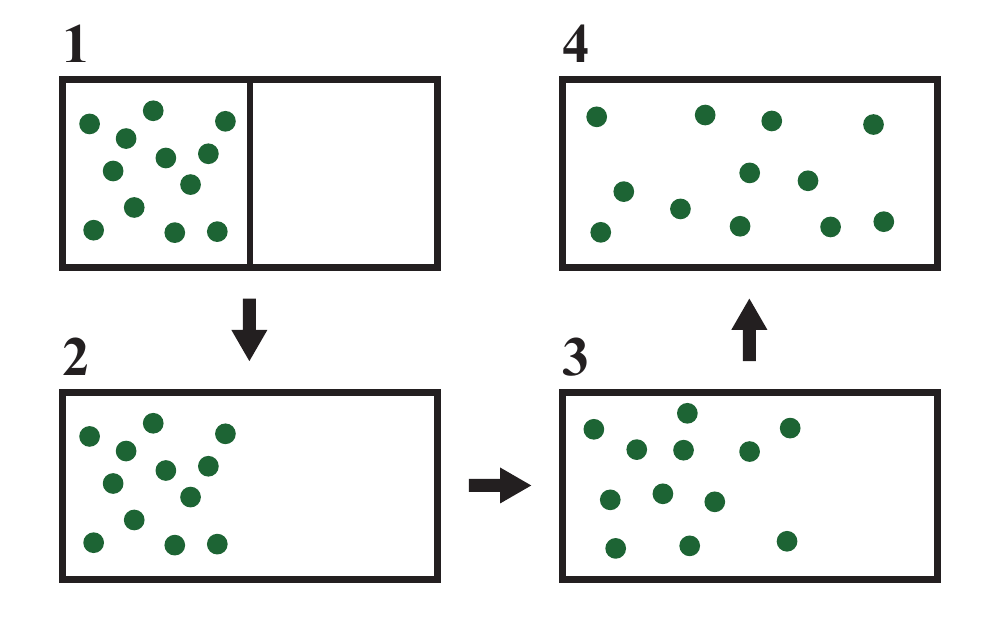}
\ \ \ \ \ \ \ \ (a)
\includegraphics[width=1\hsize]{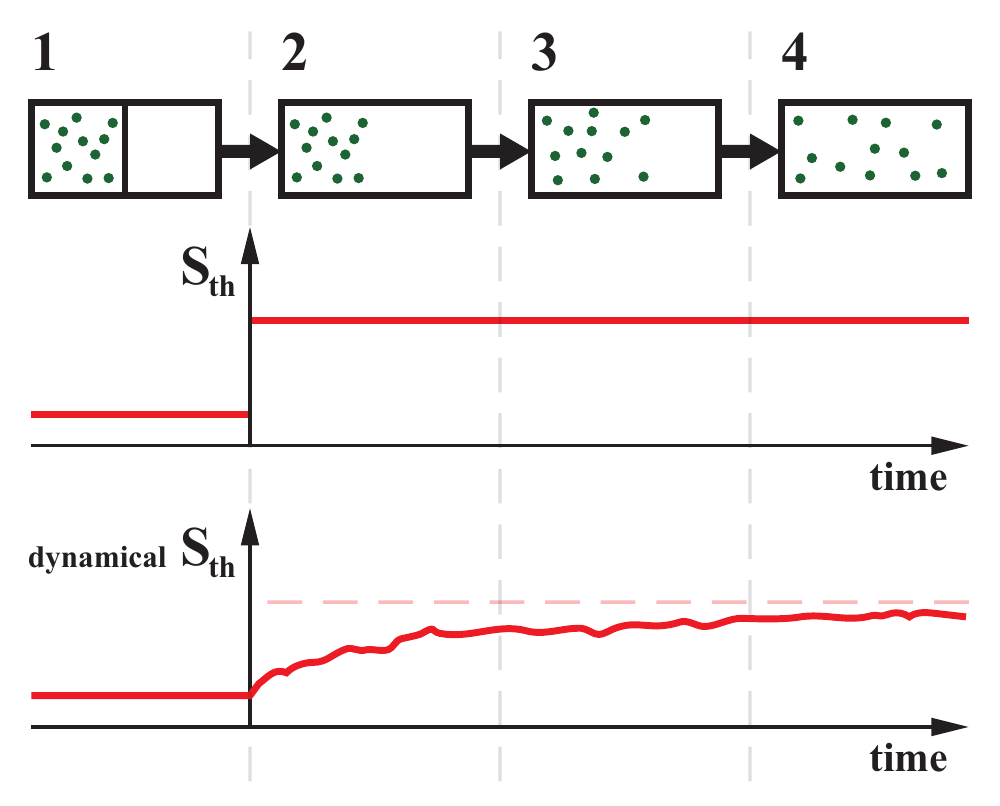}
\ \ \ \ \ \ \ \ (b)
\caption
{
(a) Evolution of an expanding gas. (b) Thermodynamic entropy $S_{\mathrm{th}}$, calculated as a logarithm of energy density of states, discontinuously increases from $1 \rightarrow 2$, because the Hamiltonian (or equivalently, boundary conditions) discontinuously changes. Then it stays constant. One motivation of this paper is to find an entropy measure that describes the \emph{dynamical} process of equilibration, i.e., a measure that depends on the state of the system rather than on the boundary conditions. Such measure is expected to stay constant as $1 \rightarrow 2$, to increase during $2 \rightarrow 4$, and to be approximately equal to thermodynamic entropy at points $1$ and $4$, when the system is in equilibrium. We would call such measure a \emph{dynamical thermodynamic entropy}.
}
\label{Fig:expansionscheme}
\end{center}
\end{figure}

\begin{figure}[t]
\begin{center}
\includegraphics[width=1\hsize]{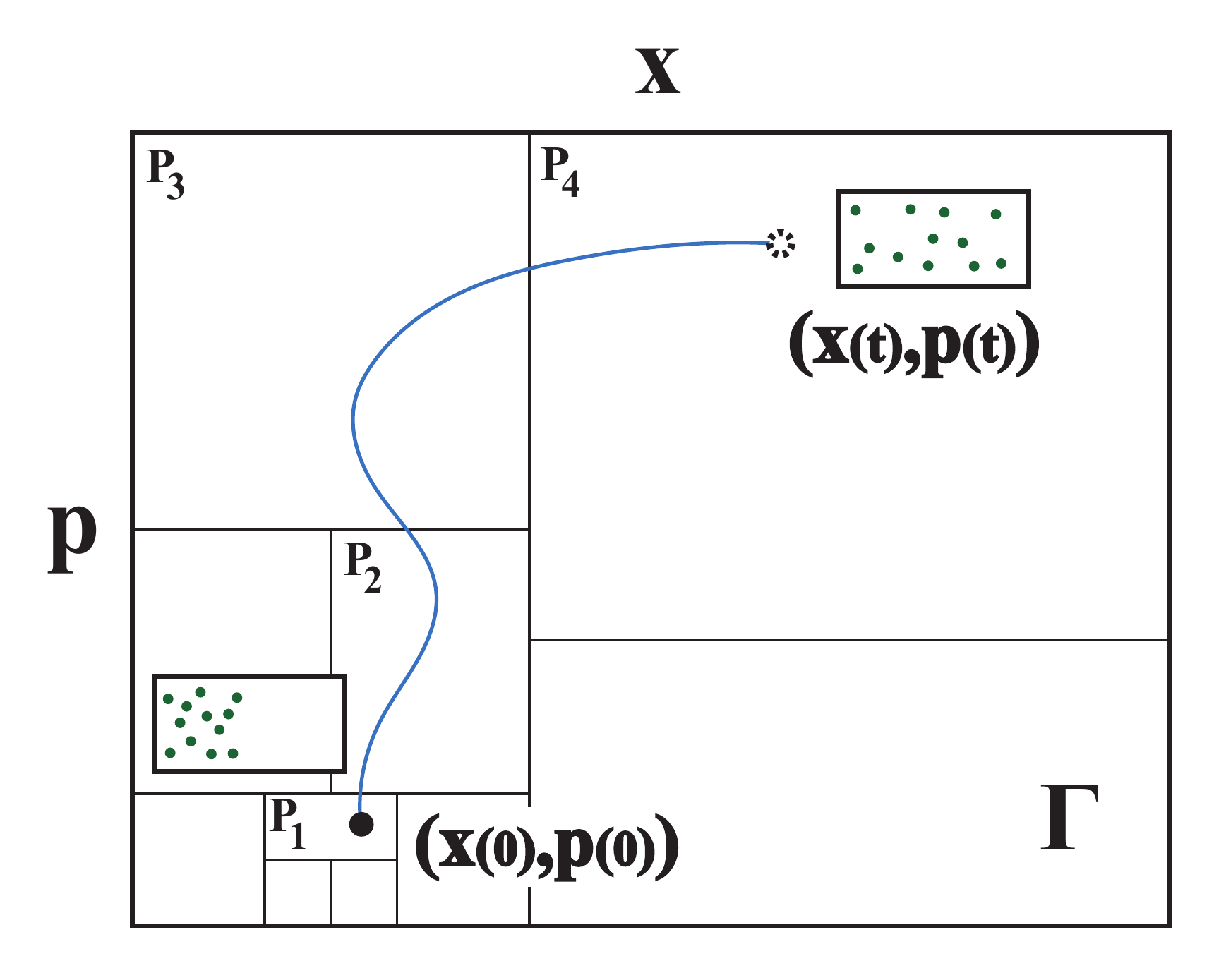}\\
\includegraphics[width=1\hsize]{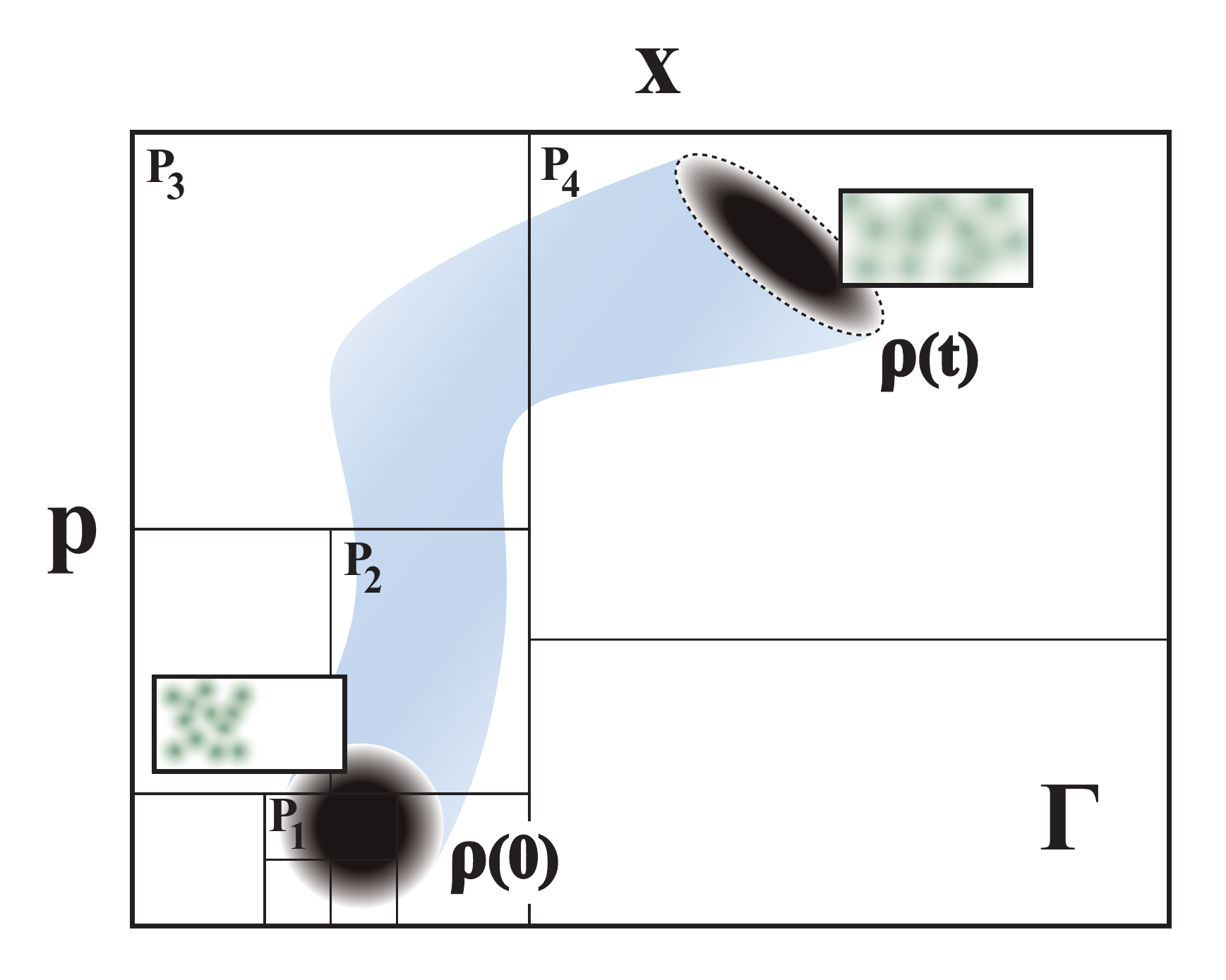}
\caption
{
Schematic picture of an evolution of a system through phase-space. The point in phase-space naturally evolves from a small macrostate $P_1$ to a large macrostate $P_4$ and spends most of its time there. The second picture shows the same evolution, but with indeterminate initial conditions signifying that the exact positions and momenta of particles are not known. In such situation the state of the system can span over several macrostates at the same time.
}
\label{Fig:evolutioninphasespace}
\end{center}
\end{figure}

This entropy typically rises, at least on average, in any classical dynamical system out of equilibrium. Consider for example the prototypical system of a small box of gas that is opened within a larger box, as depicted in Fig.~\ref{Fig:expansionscheme}, panel (a). In the Boltzmann view, we consider the phase-space of gas particles in the full box, so that immediately post-opening, the gas is in a low-entropy macrostate that might be described as ``all particles in the small box."  Under natural evolution the microstate tends to wander out of this macrostate and into the much larger macrostate ``particles spread throughout large box.'' This is depicted in Fig.~\ref{Fig:evolutioninphasespace} (top panel.)  

This scheme has the advantage of being defined {\em throughout} the process, not just in the equilibrium states.  But it is problematic in that it changes discontinuously as the microstate transitions from one macrostate to another, and in that it requires perfect knowledge of the microstate, which is never in practice available. 
What if one would like to describe a more realistic situation, in which the observer has only a partial knowledge about the state of the system? We will show that in such situations, classical Observational entropy extends the definition of Boltzmann, and leads to the continuously evolving quantity.

Another goal of this paper, as previously mentioned, is to use this generalized and smoothly-varying definition of Boltzmann entropy to describe thermodynamic entropy as a {\em dynamical} quantity, as opposed to the standard definition, which yields a fixed value that is completely determined by the external parameters of a system. To elaborate, in equilibrium systems, thermodynamic entropy (or more precisely, microcanonical entropy) is defined in terms of the density of states near a given overall conserved energy. Illustrated on our example, thermodynamic entropy defined this way changes only during the sudden quench of the Hamiltonian, which changes discontinuously due to the removal of the barrier during stage $1 \rightarrow 2$ in Fig.~\ref{Fig:expansionscheme} (a). In other words, this prescription ascribes exactly the same entropy to states 2 and 4, as illustrated in Fig.~\ref{Fig:expansionscheme} (b) (top).  This conflicts with the fact that the state 2 is a highly non-equilibrium state from which it should be in principle possible to extract a larger amount of work than from the equilibrium state 4. 
We will show that with a suitable choice of coarse-graining it is indeed possible to define a thermodynamic entropy that does not depend on the barrier removal, but rather on the underlying state of a system, as shown of Fig.~\ref{Fig:expansionscheme} (b) (bottom). Moreover, this quantity will correspond to the standard thermodynamic entropy in its respective equilibrium states, increasing continuously from one equilibrium configuration (closed box) to the next (open box after a long time), making it a reasonable definition of a dynamical thermodynamic entropy.

\section{Classical Observational entropy}
We desire a generalization of Boltzmann entropy that can be applied to probability distributions given by phase-space density
\[
\rho(\bx,\bp;t)
\]
over microstates. This dynamical entropy should act as smoothing-out of the Boltzmann entropy, reducing to the actual Boltzmann entropy for a fully determined system. We also choose the phase-space density to be normalized as
\[\label{eq:normalization_of_rho}
\int_{\Gamma} \rho(\bx,\bp;t) d\bx d\bp=1.
\]
By phase-space density $\rho(t):(\bx,\bp)\rightarrow \rho(\bx,\bp;t)$ we mean a function parametrized by time $t$ which attaches a non-negative-valued probability density to each point in phase-space.
Thus the probability that the state of the system is in the infinitesimal phase-space volume $d\bx d\bp$ at time $t$ is given by $\rho(\bx,\bp;t) d\bx d\bp$.

In the case of indistinguishable particles, the phase-space density is additionally required to be fully symmetric under the interchange of any two particles.

We assume that an observer -- guided by some physical motivation -- chooses a certain partition of this space, $\Gamma=\bigcup_iP_i$. We collect these disjoint subsets -- regions of phase-space -- into a what we call a \emph{coarse-graining}, and denote $\C=\{P_i\}_i$. 

Each region $P_i$, called a \emph{macrostate}, is commonly defined by some inequality conditions on points of phase-space $(\bx,\bp)$. It  usually represents a collection of points of phase-space consistent with some macroscopically observed value or property. 

The number of states in each macrostate, which we simply call its \emph{volume}, is linearly proportional to its phase-space volume. The volume of each macrostate is computed as an integral over the macrostate with a constant\footnote{We demand the constant measure for the following reason: Due to Liouville's evolution, the phase-space volume of any state described by a phase-space density $\rho$ is constant. If measure $\mu$ varied depending on $(\bx,\bp)$, then the volume $V_\rho=\int_{\{(\bx,\bp)|\rho(\bx,\bp;t)\neq 0\}} \mu d\bx d\bp$ associated with phase-space density would change in time. Since this volume is supposed to represent the number of microstates within it, and since it is reasonable to demand that this number stay the same during the evolution, we must demand a constant measure $\mu$. More generally, Liouville's equation and the property of time independent phase-space volume is one of the primary reasons for using phase-space to describe thermodynamics, as opposed to for example configuration space, where the volume of states does not stay constant during time evolution.} measure $\mu$,
\[\label{eq:volume}
V_i=\int_{P_\bi} \mu\  d\bx d\bp.
\]
This constant measure determines the inverse of the size of each microstate, and we will typically choose a conventional
\[
\mu=\frac{1}{h^{Nd}},
\]
where $d$ is number of degrees of freedom of each particle ($d=1$ for a particle moving on a line, and $d=3$ for particle in a 3-dimensional space), $N$ is number of particles, and $h$ is Planck's constant. The advantage of this specific choice is that its dimensions render $V_i$ unitless (as required to take a logarithm), and because
$V_i$ then measures the actual number of quantum microstates within each macrostate, since each quantum microstate is considered to take a phase-space volume of $h^{Nd}$, see e.g.~\cite{balian2007microphysics232,goodstein2014states}.

In the case of indistinguishable particles, the microstate where the $k^{\mathrm{th}}$ particle is at position and momenta $(x,p)$ and the microstate where $l^{\mathrm{th}}$ particle is at the same point in phase space, are considered to be the same microstate. As a result, Eq.~\eqref{eq:volume} is overcounting the number of microstates, and has to be modified by a standard factor~\cite{balian2007microphysics} to $V_i=\frac{1}{N!}\int_{P_\bi} \mu\  d\bx d\bp.$

To calculate the \emph{probability of the state being in macrostate $P_i$}, we integrate the phase-space density over this region,
\[
p_i(t)=\int_{P_i} \rho(\bx,\bp;t) d\bx d\bp.
\]

This gives a positive value, and all probabilities sum up to one, since by definition, integral over the entire phase-space is equal to one by Eq.~\eqref{eq:normalization_of_rho}.

Defining \emph{projectors} as window functions
\[\label{eq:window_func}
\P_i (\bx,\bp)=\begin{cases}
      1, & (\bx,\bp)\in P_i \\
      0, & (\bx,\bp)\not\in P_i
   \end{cases}
\]
we can  write
\begin{subequations}
\begin{align}
V_i&=\mu\int_{\Gamma} \P_i (\bx,\bp)  d\bx d\bp.\\
p_i(t)&=\int_{\Gamma} \rho(\bx,\bp;t) \P_i (\bx,\bp) d\bx d\bp,
\end{align}
\end{subequations}
where the integral now goes over the entire phase-space. This can be further compacted using the $L^2$-inner product on real-valued functions,
\[\label{eq:inner_product}
(f,g)\equiv \int_{\Gamma} f(\bx,\bp) g(\bx,\bp) d\bx d\bp,
\]
so that
\begin{subequations}\label{eq:V_and_p}
\begin{align}
V_i&= (\P_i,\mu),\\
p_i(t)&= (\P_i,\rho(t)).
\end{align}
\end{subequations}
($V_i=\frac{1}{N!}(\P_i,\mu)$ for indistinguishable particles.)

Since $V_i$ denotes the number of microstates in macrostate $i$, and $p_i$ denotes the probability of being in that macrostate, then assuming an observer cannot distinguish between any two microstates within the same macrostate, they associate equal probability $p_i^{(k)}:=p_i/V_i$ to each microstate $k$ within that macrostate. The Shannon entropy that this observer assigns to a system, given their inability to distinguish microstates within the same macrostate, is therefore 
$-\sum_{i,k} p_i^{(k)}\ln p_i^{(k)}=-\sum_{i}V_i\frac{p_i}{V_i}\ln \frac{p_i}{V_i}$. This motivates the definintion of {\em classical Observational entropy with coarse-graining} $\C$ as
\[\label{eq:definition_of_OE}
S_{O(\C)}(t)=-\sum_ip_i(t)\ln\frac{p_i(t)}{V_i}.
\]
Since the coarse-graining $\C$ can be uniquely defined either by a set of macrostates $P_i$, or by a set of corresponding projectors $\P_i$, we can identify these two otherwise mathematically distinct objects, and write
\[
\C=\{P_i\}_i\equiv\{\P_i\}_i.
\]

So far, our definition of coarse-grained entropy is mathematically identical to those used by Refs.~\cite{wehrl1978general,nauenberg2004evolution,kozlov2007fine,piftankin2008gibbs,vzupanovic2018relation}; the next section introduces an important generalization to multiple coarse-grainings.

\section{Multiple coarse-grainings}

The above definition applies for any coarse-graining, but some coarse-grainings are more relevant than others.
To define a thermodynamically relevant specification of Observational entropy, and for other purposes, it is necessary to generalize this entropy to include multiple coarse-grainings. This is done as follows.

Suppose we have several coarse-grainings of the phase-space, $(\C_1,\dots,\C_n)$. We define a \emph{joint coarse-graining}
\[
\C_{1,\dots,n}=\{P_{i_1,\dots,{i_n}}\}_{i_1,\dots,{i_n}}
\]
where the corresponding multi-macrostates are given by overlap of the previous macrostates, and the projectors as a multiple of projectors,
\begin{align}
P_\bi&\equiv P_{i_1,\dots,{i_n}}\equiv P_{i_1}\cap\cdots\cap P_{i_n},\\
\P_\bi&\equiv \P_{i_1,\dots,{i_n}}\equiv \P_{i_1}\cdots \P_{i_n}.
\end{align}
In the above, we have also employed multi-index ${\bi=(i_1,\dots,i_n)}$.

Inserting the joint coarse-graining into the definition, Eq.~\eqref{eq:definition_of_OE}, motivates the definition of the Observational entropy with multiple coarse-grainings as
\[\label{eq:definition_of_generalOE}
S_{O(\C_1,\dots,\C_n)}(t)\equiv-\sum_\bi p_\bi(t)\ln\frac{p_\bi(t)}{V_\bi},
\]
where
\begin{align}
V_\bi&= (\P_\bi,\mu),\\
p_\bi(t)&= (\P_\bi,\rho(t)).
\end{align}
($V_\bi=\frac{1}{N!}(\P_\bi,\mu)$ for indistinguishable particles.) Indeed, from the definition it follows $S_{O(\C_1,\dots,\C_n)}=S_{O(\C_{1,\dots,n})}$. (Note the subtle notational difference.)
That is, a set of coarse-grainings can also be considered as a single composite coarse-graining.\footnote{As discussed below, this statement does not transfer to quantum mechanics unless the coarse-grainings commute.}

\section{Properties}\label{sec:properties}

We briefly mention some properties of classical Observational entropy. These are classical equivalents of theorems proved for the quantum Observational entropy~\cite{safranek2019long}. The proofs are mostly analogous to those quantum mechanics (with some important subtleties), although we were able to slightly simplify some of them due to a simpler structure of the phase-space as compared to the Hilbert space (see~Appendix~\ref{app:proofs}).

\begin{theorem}\label{thm:Boltzman_generalization} (Observational entropy is a generalization of the Boltzmann entropy)
For a single point in phase-space $(\bx,\bp)\in P_i$, equivalent to a delta function $\rho(\tilde\bx,\tilde\bp)=\delta(\tilde\bx-\bx,\tilde\bp-\bp)$, we have
\[
S_{O(\C)}(\rho)=S_B(\rho)=\ln V_i.
\]
\end{theorem}

\begin{definition}\label{def:finer_coarse_graining} (Finer coarse-graining)
We say that coarse-graining $\C_2$ is {\em finer} than coarse-graining $\C_1$ (and denote $\C_1\hookrightarrow \C_2$) when for every $P_{i_1}\in\C_1$ there exists an index set $I^{(i_1)}$ such that $P_{i_1}=\bigcup_{i_2\in I^{(i_1)}}P_{i_2}$, ${P}_{i_2}\in \C_2$. (That is, each element of $\C_1$ can be partitioned using elements of $\C_2$.)
\end{definition}

When $\C_1\hookrightarrow \C_2$, we can also write $\P_{i_1}=\sum_{i_2\in I^{(i_1)}}\P_{i_2}$.

\begin{theorem}\label{thm:monotonic} (Observational entropy is a monotonic function of the coarse-graining.)
If $\C_1\hookrightarrow \C_2$ then
\[
S_{O(\C_1)}(\rho)\geq S_{O(\C_2)}(\rho).
\]
\end{theorem}

\begin{definition}\label{def:observablecg} (Coarse-graining given by an observable) Let $A:(\bx,\bp)\rightarrow a$ be a classical observable, that assigns value $a$ (property) to each point in phase-space. We define macrostates associated with value $a$ as ${P_a=\{(\bx,\bp)|A(\bx,\bp)=a\}}$ in case of observable with discrete values, or as ${P_a=\{(\bx,\bp)|a\leq A(\bx,\bp)<a+da\}}$ in case of a continuous observable.\footnote{The infinitesimal increment $da$ plays the role of the resolution in measuring the observable $A$.} We define coarse-graining given by the observable\footnote{Note 1: Although $\rho$ is not usually considered an observable, it fulfills our definition and we will treat it as such in the theorem that follows. Note 2: we can also define the spectral decomposition of an observable as $A=\sum_a a\P_a$, or $A=\int_{-\infty}^\infty a \P_a da$, where eigenvalues $a$ are considered to be different from each other. This spectral decomposition is unique, and it plays an identical role in phase-space with inner product~\eqref{eq:inner_product} as the spectral decomposition of a quantum observable plays in a Hilbert space.} $A$ as $\C_A=\{P_a\}_a$. 
\end{definition}

\begin{theorem}\label{thm:bounded_multiple} (Observational entropy with multiple coarse-grainings is bounded)
\[
S_G(\rho)\leq S_{O(\C_1,\dots,\C_n)}(\rho)\leq \ln V
\]
where $S_{G}(\rho)\equiv-\int_\Gamma\frac{\rho(\bx,\bp)}{\mu}\ln \frac{\rho(\bx,\bp)}{\mu}\ \mu d\bx d\bp$, and $V\equiv\int_\Gamma \mu d\bx d\bp$ is the total volume of the phase-space. $S_G(\rho)= S_{O(\C_1,\dots,\C_n)}(\rho)$ if and only if $\C_\rho\hookrightarrow \C_{1,\dots,n}$, i.e., if the joint coarse-graining is fine enough to distinguish between points in phase-space that have  different assigned probabilities.
\end{theorem}

$S_G(\rho)$ represents Gibbs entropy.\footnote{Definition of Gibbs entropy is not consistent in literature. Our definition seems to be the most common, and is used for example in~\cite{jaynes1965gibbs}. Sometimes, Gibbs entropy is defined as Shannon entropy of probabilities of energy distribution (which has a quantum equivalent called diagonal entropy~\cite{polkovnikov2011microscopic}), and what is called Gibbs entropy in~\cite{vzupanovic2018relation} is actually our Observational entropy.} Gibbs entropy is invariant under Liouville's evolution. It is zero for a single point in phase-space, and it is a property of a state, and not of a coarse-graining. This quantity also appears as the functional $H$ in the classical H-theorem, as interpreted by Tolman~\cite{tolman1979principles}. Since $\mu$ is the inverse of the phase-space volume of a single microstate, $p_i=\frac{\rho(\bx,\bp)}{\mu}$ is the probability of being in a microstate, so we can as well write $S_{G}(\rho)\equiv-\sum_ip_i\ln p_i$, i.e. the Shannon entropy of these probabilities. The quantum mechanical equivalent of Gibbs entropy is the von Neumann entropy, which is invariant under unitary evolution, is zero for pure states, and which is also property of a state and not of a coarse-graining.

\begin{theorem}\label{thm:non-increase} (Observational entropy is non-increasing with each added coarse-graining.)
\[
S_{O(\C_1,\dots,\C_n)}(\rho)\leq S_{O(\C_1,\dots,\C_{n-1})}(\rho)
\]
for any set of coarse-grainings $(\C_1,\dots,\C_n)$ and any phase-space density $\rho$.
\end{theorem}

These theorems show that Observational entropy can be elegantly interpreted as the amount of knowledge an observer would obtain if he or she were to measure the macroscopic observables that define the coarse-grainings. While Theorem~\ref{thm:monotonic} says that an observer with better resolution will get to know more about the system, Theorem~\ref{thm:bounded_multiple} says that no matter what coarse-grained measurements they choose to perform, their knowledge will be still limited by an inherent uncertainty in the system given by the Gibbs entropy. On the other hand, no matter the coarse-graining, their knowledge cannot be worse than that measured by the maximal entropy, which signifies complete uncertainty about the system's state. Theorem~\ref{thm:non-increase} then shows an intuitive statement that each additional macroscopic measurement will provide better knowledge of the system, at least on average.

In addition, rewriting Eq.~\eqref{eq:definition_of_generalOE} as
\[
S_{O(\C_1,\dots,\C_n)}(t)\equiv-\sum_\bi p_\bi(t)\ln p_\bi(t)+\sum_\bi p_\bi(t)\ln V_\bi
\]
provides an intuitive information-theoretic interpretation. The first term denotes an uncertainty as measured by the Shannon entropy regarding to which macrostate the microstate of the system belongs. In other words, if one were to make a coarse-grained measurement at time $t$ given by the coarse-graining (for example a measurement determining the system's energy) the first term measures the uncertainty in the measurement outcomes in such a coarse-grained measurement. The second term measures the average remaining uncertainty about the microstate after this coarse-grained measurement was done. Put together, Observational entropy measures the average amount of uncertainty about a microstate of a system, from a point of view of an observer that can track only certain macroscopic properties by his/her ability to perform coarse-grained measurement.

\section{Thermodynamically relevant non-equilibrium entropies: introduction}

The treatment thus far has pertained to any possible coarse-graining. However, even if entropy increase is generic, there is no reason to expect that an arbitrary coarse-graining will be closely connected with {\em thermodynamics}, which in particular relates temperature, energy and entropy.

In the following sections we introduce two types of (composite) coarse-grainings that each lead to an entropy that is relevant for describing dynamics of isolated quantum systems: $S_{xE}$ (Observational entropy with local particle coarse-graining and global energy coarse-grainings) and $S_F$ (Observational entropy with local particle and local energy coarse-grainings).\footnote{In the quantum context~\cite{safranek2019long}, $S_F$ is also named ``Factorized Observational entropy'' or FOE for short, due its tensor-product form of its local energy coarse-graining.} Both will be perfectly defined for systems outside of equilibrium, both will grow to the equilibrium thermodynamic value when the system thermalizes, and both will describe physical regions coming into equilibrium with each other.

However, there are some important distinctions in their properties closely related to distinctions in their meanings: as explained in the following sections, $S_{xE}$ can be interpreted as the value of entropy the system would attain in the long-time limit if the regions were allowed to exchange energy, but not particles. $S_F$, on the other hand, represents the value of entropy associated with the system if regions equilibrate but cannot exchange either energy or particles.

This interpretation of $S_{xE}$ shows that it is not in general additive on subsystems. For example, given two equally-sized regions consisting of the same number of particles but with very different temperature, $S_{xE}$ assigns an entropy value appropriate to the two systems' temperatures having already equalized through interaction between the regions. $S_F$, on the other hand, is additive, and equal to the sum of the thermodynamic entropies of the two regions.  Additionally, since measuring local energies also determines the global value of energy, in classical physics $S_F \le S_{xE}$ always.

Since in real systems both particles and energy are exchanged between the regions, both $S_{xE}$ and $S_F$ are inherently time-dependent quantities outside of equilibrium. 

In the following Sections, we focus on $S_{xE}$ (Sections~\ref{sec:sxeDef} and~\ref{sec:thermalsxe}), not because it is of more interest but because its definition requires only the global Hamiltonian, making it more transparent. We will derive its dynamical properties and show a simulation illustrating its behavior. Defining $S_F$ requires a notion of local Hamiltonians. Despite this, its properties are derived in a close analogy to those of $S_{xE}$ (Section~\ref{sec:SF}).\footnote{
While both definitions can be written in the standard Hamiltonian formalism which assumes there is a fixed number of particle in the entire system, the description is much more elegant when one develops a formalism of the classical analogue of Fock space -- the \emph{Fock phase-space} -- which allows for describing classical systems with a variable number of particles. After all, each subsystem (both for $S_{xE}$ and $S_F$) has a variable number of particles in it, and therefore it should be possible to describe each subsystem independently. Observables of local particle number and local energies will be then seen as observables on the Fock phase-space of a subsystem. The coarse-grainings defining $S_F$ is then created as a Cartesian product of local coarse-grainings. This will be done elsewhere~\cite{safranek2020classicalee}.
}

Both of these entropies, $S_{xE}$ and $S_F$, originated as fully quantum mechanical versions published in~\cite{safranek2019letter,safranek2019long}. However, there are some important differences. For example, unlike classical $S_{xE}$, the quantum $S_{xE}$ is (quite surprisingly!) an additive quantity. These will be discussed in detail in Sec.~\ref{sec:class_quant}.

\section{$S_{xE}$: entropy of measuring local particle numbers and global energy}\label{sec:sxeDef}

In this section we introduce a (composite) coarse-graining that defines one of the thermodynamically relevant entropies, $S_{xE}$.

Let us consider a system of $N$ particles contained in a 1-dimensional box from position $L_1$ to $L_2$, so of size $L=L_2-L_1$. We coarse-grain this box into $m$ physical regions (\emph{bins}) of size $\Delta x=\frac{L}{m}$. Considering the vector of positions as $\bx=(x_1,\dots,x_N)$, and vector of number of particles in each part of the box as $\bn=(n_1,\dots,n_m)$ (where $n_1+\cdots+n_m=N$)
we define \emph{local particle number (configuration) macrostates} as
\[
\begin{split}
&P_\bn\equiv \{(\bx,\bp)|n_1 \text{ particles with position } x \in [L_1,L_1\!+\!\Delta x),\\
&\ \dots, n_m \text{ particles with position } x \in [L_1\!+\!(m\!-\!1)\Delta x,L_2)\}.
\end{split}
\]
Clearly, we can easily generalize this to any number $d$ of spatial dimensions. This coarse-graining corresponds to measuring number of particles in each one of the $m$ bins, i.e., to a coarse-grained measurement of local particle numbers, and we denote it $\C_X=\{P_\bn\}_\bn$.\footnote{We denote coarse-graining $\C_X$ with lower index $X$ signifying position mostly for historical reasons, because in the case of indistinguishable particles focused on in the quantum version of observational entropy~\cite{safranek2019letter,safranek2019long}, measuring coarse-grained positions of particles is the same as measuring the local number of particles. Thus also an equivalent name for $\C_X$ is positional coarse-graining.} For example, $P_{(3,0,1)}$ corresponds to a macrostate where three particles are in the first bin, zero in the second, and one in the third.

We also define \emph{energy macrostates} with width $\Delta E$, as
\[
P_{E,\Delta E}\equiv \{(\bx,\bp)|E\leq H(\bx,\bp)<E+\Delta E\},
\]
where $H$ denotes the Hamiltonian. When $\Delta E\equiv dE$ is an infinitesimal increment, we simply denote $P_E\equiv P_{E,dE}$, its corresponding projector as $\hat{P}_E\equiv \hat{P}_{E,dE}$, and volume as  $V_E\equiv V_{E,dE}$. We call the coarse-graining with such an infinitesimal energy increment a ``fine-graining'' in energy, and denote it $\C_E=\{P_E\}_E$ (considering definition~\eqref{def:observablecg}, $\C_E\equiv \C_H$). It is important to emphasize here that $dE$ is fixed to be the same for all energies $E$.\footnote{For example, one could also think of an energy coarse-graining $C_E^{(\mathrm{eigen})}=\{P_E^{(\mathrm{eigen})}\}_E$, consisting of projectors onto ``eigenstates'' of the classical Hamiltonian, where we fit the width of each energy shell $dE$ so that $V_E^{(\mathrm{eigen})}=1$, meaning that each macrostate corresponds to a single microstate. In that case $dE=dE(E)$ will generally depend on the energy. Projectors (window functions) $\P_E^{(\mathrm{eigen})}$ are then a classical equivalent of
projectors onto quantum energy eigenstates $\pro{E}{E}$. However, this would not give a desirable long-time behavior for $S_{xE}$; in particular, the entropy assigned to a microstate in a given energy macrostate $P_E$ would not be the correct microcanonical entropy, which is defined as a logarithm of number of microstates in energy shell $[E,E+dE)$. Projectors $\P_E\in\C_E$ with fixed $dE$ correspond to the sum $\sum_{E\leq\tilde{E}<E+dE}\pro{\tilde{E}}{\tilde{E}}$ in quantum mechanics. We also note that correspondence between projectors onto eigenstates of the Hamiltonian $\pro{E}{E}$ (called \emph{stationary Liouville eigenstates}) and $\hat{P}_E^{(\mathrm{eigen})}$ have been explored in~\cite{wilkie1997quantum} and references therein. It was shown that the Wigner function of $\pro{E}{E}$ converges to a distribution that can be viewed as a highly peaked $\hat{P}_E^{(\mathrm{eigen})}$ in the $h\rightarrow 0$ limit (see Eq.~(24) in~\cite{wilkie1997quantum}).\label{footnoteE}}

\begin{figure}[t]
\begin{center}
\includegraphics[width=1\hsize]{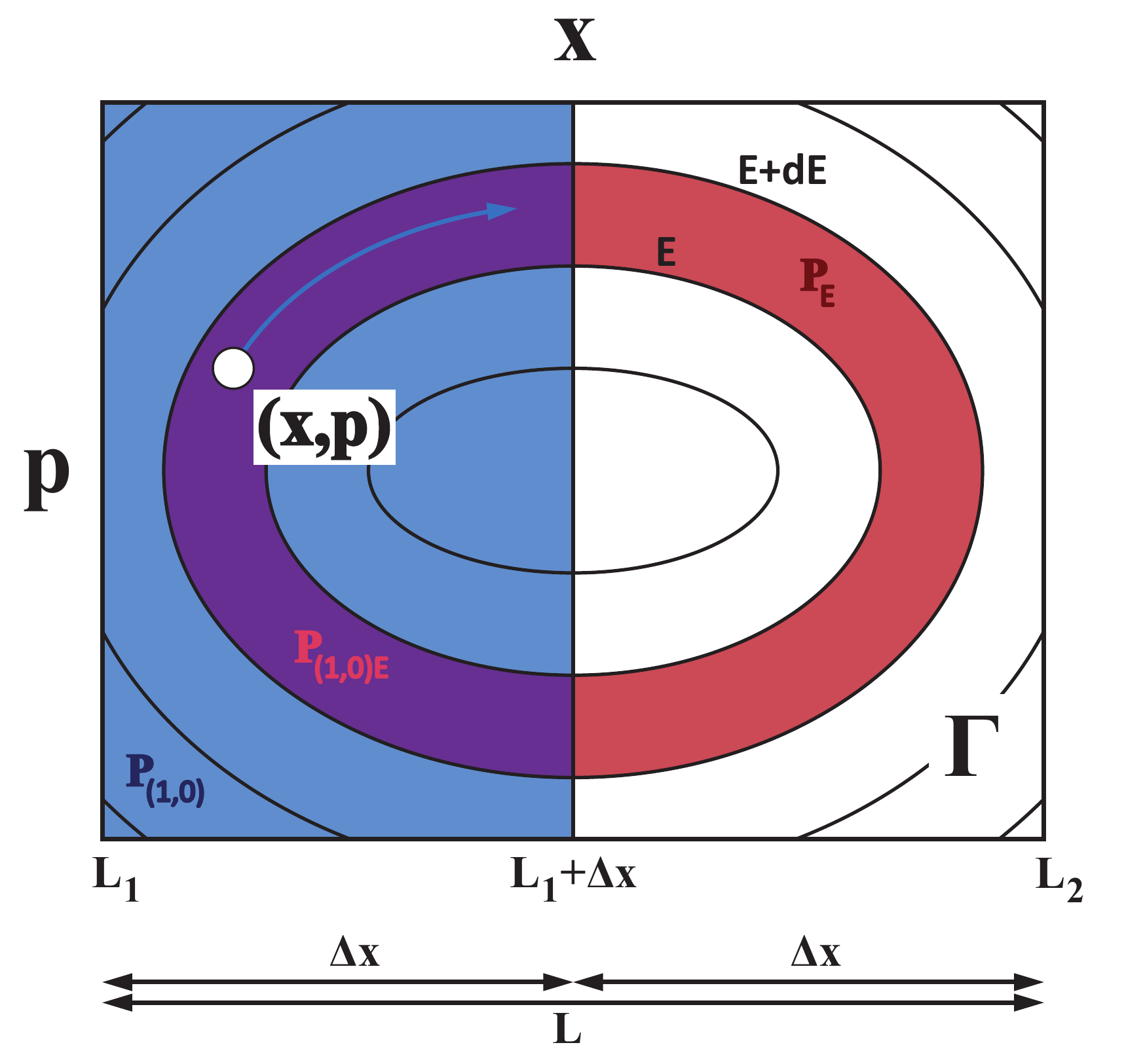}
\caption
{Phase-space of a single particle in a Harmonic potential, with a local particle number (configuration) coarse-graining, and energy coarse-graining. The blue local particle number macrostate $P_{(1,0)}$ corresponds to statement ``the particle is on the left side of the box'', while the red energy macrostate $P_E$ corresponds to ``particle has energy between $E$ and $E+dE$''. Overlap of these two macrostates $P_{(1,0)E}$ (purple) is a local particle number-energy macrostate which corresponds to ``the particle is on the left side of the box and has energy between $E$ and $E+dE$''. Energy macrostates --- shells of constant energy of width $dE$ --- form ellipses in this example because they are given by equations $E=\frac{1}{2m} p^2+\frac{1}{2}k x^2=\mathrm{const.}$ The volume of the energy macrostate $V_E$ is proportional to the red area (including purple) in the picture, and it defines the microcanonical entropy $S_{\mathrm{micro}}(E)\equiv \ln V_{E}.$ In an isolated system, a particle will never jump out of its energy shell, but only rotate through it (blue arrow). In the situation depicted, where the volume of the macrostate with particle on the left is the same as the volume of macrostate with particle on the right, $V_{(1,0)E}=V_{(0,1)E}=\frac{1}{2}V_{E}$, the dynamical thermodynamic entropy has a constant value $S_{xE}(t)=\ln(\frac{1}{2}V_{E})=S_{\mathrm{micro}}(E)-\ln 2$, which is the same as microcanonical entropy of a particle in a half of the box.
}
\label{Fig:ellipse}
\end{center}
\end{figure}

 The \emph{local particle number-energy macrostates} are then defined as ${P_{\bn E}\equiv P_{\bn} \cap P_E}$, and the respective projectors are ${\P_{\bn E}=\P_{\bn}\P_{E}}$. With macrostate volumes $V_{\bn E}= (\P_{\bn E},\mu)$, and probabilities $p_{\bn E}= (\P_{\bn E},\rho_t)$, we define observational entropy corresponding to measuring local particle numbers and the global energy as
\[\label{eq:Sxe_definition}
S_{xE}(t)\equiv S_{O(\C_X,\C_H)}(t)=-\sum_{\bn, E}p_{\bn E}(t)\ln\frac{p_{\bn E}(t)}{V_{\bn E}}.
\]
Figure~\ref{Fig:ellipse} gives an example of these quantities for a single particle in a Harmonic potential.

\begin{figure*}[hp]
\begin{center}
\includegraphics[width=1\hsize]{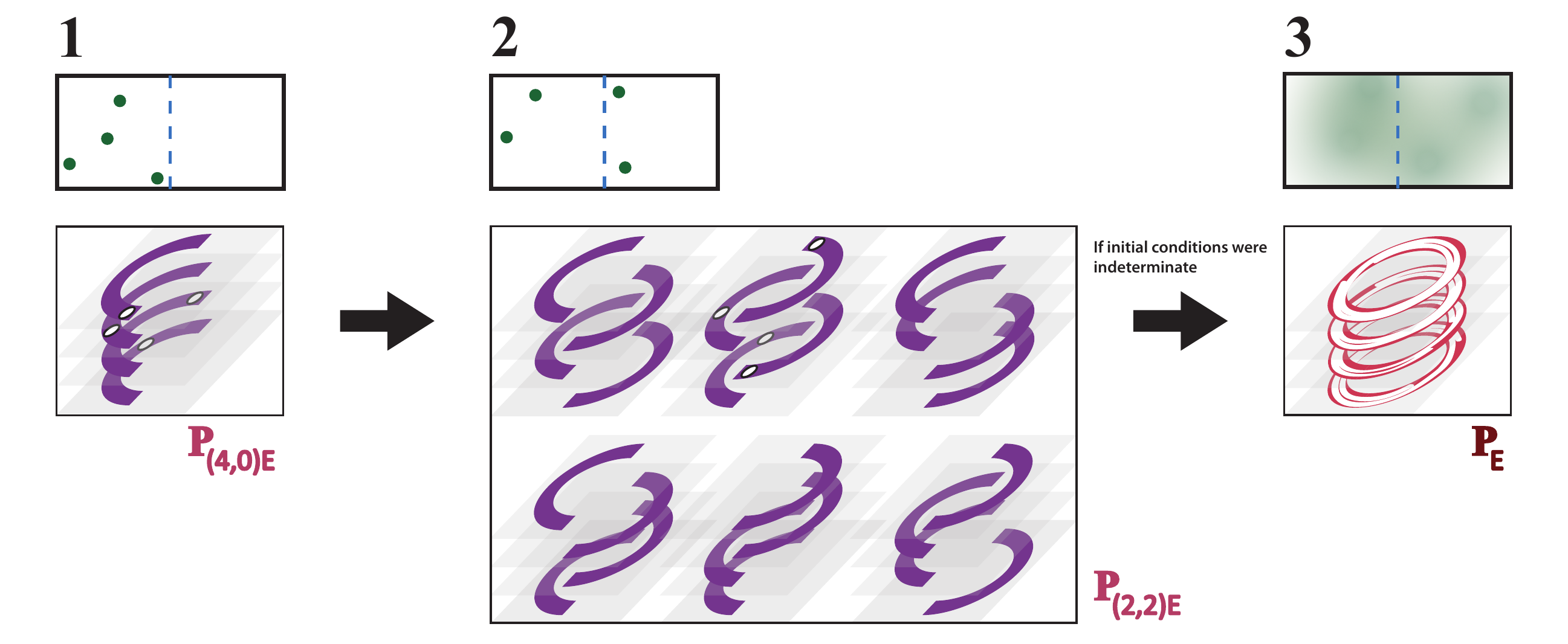}
\caption
{Evolution of N=4 indistinguishable particles in the phase-space, with local particle number-energy coarse-graining taken from Fig.~\ref{Fig:ellipse}. We plot the 16-dimensional phase-space as four single particle phase-spaces stacked onto each other. $\boldsymbol{(1)}$ Particles start in the left side of the box, which corresponds to macrostate $P_{(4,0)E}$ (``four particles on the left, zero on the right''), where $E$ denotes the total energy of the system. This configuration gives microcanonical entropy of the left half of the box, $S_{xE}=\ln V_{(4,0)E}=S_{\mathrm{micro}}^{(H_1)}(E)$. $\boldsymbol{(2)}$ As particles evolve, they wander into the largest macrostate $P_{(2,2)E}$ (``two particles on the left, two on the right'') allowed by the given energy, which is $6$ times bigger than the initial macrostate, and they spend most of their time there. The entropy of this largest macrostate is equal to the microcanonical entropy of the entire box up to some correction, $S_{xE}=\ln V_{(2,2)E}=S_{\mathrm{micro}}^{(H_2)}(E)-\frac{1}{2} \ln (2\pi)$, $m=2$ in Eq.~\eqref{eq:limittpoint}. $\boldsymbol{(3)}$ If the initial state was not fully determined, then after some time the particle positions become uncertain, and the phase-space density becomes quite uniformly smeared over the entire energy shell $P_E=\bigcup_{\bn, n_1+n_2=4}P_{\bn E}$, which is the  effect known as mixing. This erases corrections to the entropy, which then exactly equals the microcanonical entropy of the entire box: $S_{xE}=\ln V_E=S_{\mathrm{micro}}^{(H_2)}(E)$.}
\label{Fig:phasespaceevolution}
\includegraphics[width=1\hsize]{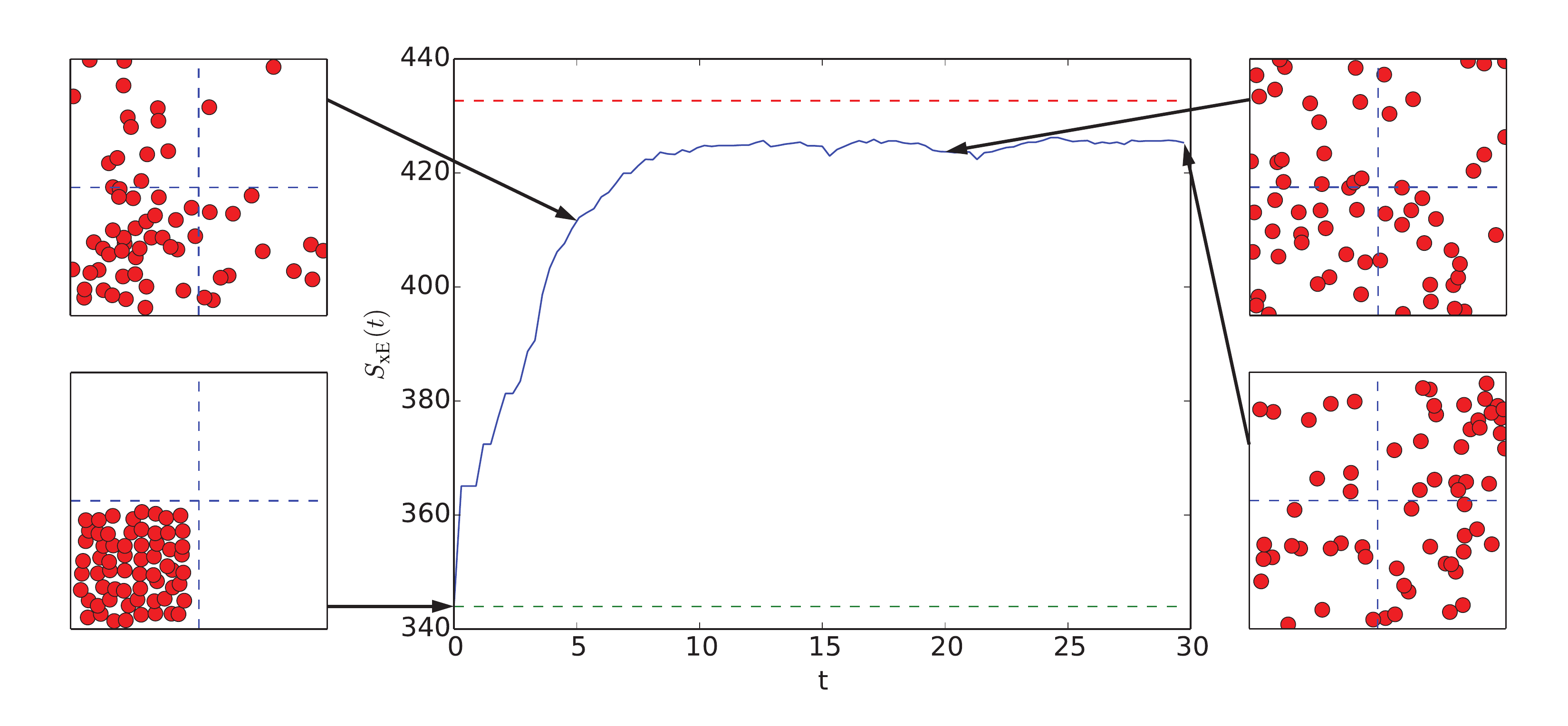}
\caption
{
Simulation of evolution of $S_{xE}$ of a system of $N=64$ particles in a 2-dimensional box coarse-grained into $m=4$ physical regions, evolving through a Hamiltonian including weak inter-particle interactions, and taking periodic boundary conditions. The green and red dashed lines represent thermodynamic entropy of the bottom left quarter, and the full system respectively. As the system evolves, the particles spread throughout the regions, and the $S_{xE}$ grows to the thermodynamic entropy of the full system, up to some finite-size corrections, as expected from Eq.~\eqref{eq:limittpoint}. Illustrations of the particle spread are the real snapshots of the system at different times of evolution in our simulation.
}
\label{Fig:evolution}
\end{center}
\end{figure*}

\section{Thermodynamic behavior of $S_{xE}$}\label{sec:thermalsxe}

We now take a closer look at the thermodynamic behavior of $S_{xE}$, using the prototypical example of a gas at first contained in the half of a box, then expanding into its full volume, as shown in Fig.~\ref{Fig:expansionscheme}. We will show that for the cases where all particles are only in one of the regions, $S_{xE}$ is equal to the thermodynamic entropy of that region, and that as the system thermalizes by letting the particles spread, $S_{xE}$ grows to the thermodynamic entropy of the full system. This entropy is therefore enough to effectively model, for example, the expansion of an ideal gas.

We model our system by a Hamiltonian undergoing quench (a sudden discontinuous change) at time $t=0$,
\[
H(\bx,\bp,t)=\begin{cases}
      H_1(\bx,\bp)=H_0(\bx,\bp)+U_1(\bx), & t<0 \\
      H_2(\bx,\bp)=H_0(\bx,\bp)+U_2(\bx), & t\geq 0
   \end{cases}
\]
where $H_0$ denotes Hamiltonian of the particles themselves (which may or may not contain interaction between the particles), and
\begin{align}
U_1(\bx)&=\begin{cases}
      0, & \forall i,\ 0<x_i<\frac{L}{2} \\
      +\infty, & \mathrm{otherwise}
   \end{cases}\\
U_2(\bx)&=\begin{cases}
      0, & \forall i,\ 0<x_i<L \\
      +\infty, & \mathrm{otherwise}
   \end{cases}
\end{align}
denote two infinite potential wells.

For Hamiltonian $H$, the microcanonical entropy is defined as the logarithm of number of microstates in each energy shell~\cite{sethna2006statistical}, which we can write in several ways as
\[\label{eq:microcanonical_entropy}
S_{\mathrm{micro}}^{(H)}(E)=\ln V_{E,\Delta E}^{(H)}=\ln (\rho(E) \Delta E)=S_{\mathrm{micro}}(E,{\cal V},N).
\]
$V_{E,\Delta E}^{(H)}$ is the volume of an energy macrostate, and $\rho(E)$ denotes the density of energy states. $\Delta E$ is the width of the energy macrostate which can be fixed to some small but non-zero value\footnote{Since $\rho(E)$ rises exponentially with $E$ and $\ln (\rho(E) \Delta E)=\ln \rho(E)+ \ln \Delta E$, the choice of $\Delta E$ ultimately does not matter, because $\ln \Delta E$ acts only as an additive constant which is small compared to the first term. Moreover, since we are usually interested in the changes in entropy, rather than in absolute values, in such situations value of this additive term becomes completely irrelevant.} that is considered to be independent of $E$. ${\cal V}$ is the physical volume occupied by $N$ particles of total energy $E$. More generally, we can define thermodynamic entropy as
\[
S_{\mathrm{th}}^{(H)}\equiv S_{O(\C_H)},
\]
which coincides with the definition of microcanonical entropy for a single point in phase-space, and with canonical entropy for a canonical distribution $\rho(\bx,\bp)=\frac{1}{Z}e^{-\beta H(\bx,\bp)}$, where $\beta$ denotes inverse temperature, and $Z=\int_\Gamma e^{-\beta H(\bx,\bp)}d\bx d\bp$ is the partition function.

In the following, we consider $S_{xE}\equiv S_{O(\C_X,\C_{H_2})}$ with a local particle number coarse-graining $\C_X$ that halves the box ($\bn=(n_1,n_2)$, $L_1=0$, $L_2=L$, $\Delta x=\frac{L}{2}$), and with energy coarse-graining $\C_{H_2}$ given by the Hamiltonian after the quench.

From Theorem~\ref{thm:non-increase}, we can immediately see that thermodynamic entropy of the entire box bounds the $S_{xE}$,\footnote{Symmetry $S_{O(\C_X,\C_{H_2})}=S_{O(\C_{H_2},\C_X)}$ holds for classical Observational entropy, but not for quantum Observational entropy~\cite{safranek2019long}.}
\[
S_{xE}= S_{O(\C_X,\C_{H_2})}=S_{O(\C_{H_2},\C_X)}\leq S_{O(\C_{H_2})}=S_{\mathrm{th}}^{(H_2)}.
\]

Now we move onto studying the actual dynamics of $S_{xE}$.

\subsection{Dynamics of $S_{xE}$ for determinate initial conditions}

First we assume a fully-determined state of $N$ particles contained in the left part of the box, described by a state $(\bx_0,\bp_0)\in P_{(N,0)E}$. We have 
\[\label{eq:initialpoint}
S_{xE}(\bx_0,\bp_0)=\ln V_{(N,0)E}=\ln V_{E}^{(H_1)}=S_{\mathrm{micro}}^{(H_1)}(E)=S.
\]
The first equality comes from Theorem~\ref{thm:Boltzman_generalization}, and the second comes from the fact that macrostate of $N$ particles in the left side of the box with energy $E$ is identical to the energy macrostate of the initial Hamiltonian, $P_{(N,0)E}=P_{E}^{(H_1)}$. Therefore, for initial particles contained in the left side of the box, $S_{xE}$ gives the microcanonical entropy of this side of the box.

As the particles evolve, some of them will go to the right side of the box and some remain. After some time, the state of the system becomes what we can describes as ``about a half of the particles are on the left, and about a half of the particles are on the right side of the box''.

We can say the same by saying that the point in phase-space will wander around, and will most likely end up in one of the largest macrostates. In the current situation, the largest macrostate corresponds to a statement ``half of the particles are on the left side of the box, half are on the right, and the total energy is $E$''. This is schematically depicted as stages $1 \rightarrow 2$ in Fig.~\ref{Fig:phasespaceevolution}.

Due to the slow growth of the logarithm, the entropy associated with either of those large macrostates will not differ much: there will be some corrections, but they will become irrelevant in the thermodynamic limit. So in order to find the long-time behavior of $S_{xE}$, it is therefore enough to calculate the entropy for the largest of the macrostates, which is what we do in Appendix~\ref{app:single_point}. Assuming indistinguishable particles, we find that for $m$  bins,\footnote{The proof of an equivalent long-time limit behavior for quantum $S_{xE}$ is very different than the present classical case; instead of computing entropy of the largest macrostate, it uses ansatz of random Hamiltonian, and assumption that in the long-time limit the phases in energy basis can be considered to be drawn randomly from a uniform ensemble (which is equivalent of saying that ratios of any two energy eigenvalues are irrational)~\cite{safranek2019long}.\label{classicalvsquantumproof}}
\[\label{eq:limittpoint}
\lim_{t\rightarrow +\infty}S_{xE}(t)=S_{\mathrm{micro}}(E,L,N)+\frac{1}{2} \ln (2\pi N) -\frac{m}{2}\ln\frac{2\pi N}{m},
\]
where $S_{\mathrm{micro}}(E,L,N)=S_{\mathrm{micro}}^{(H_2)}(E)$ denotes the total microcanonical entropy of the full system. This means that in the case of perfect knowledge of the system, $S_{xE}$ matches microcanonical entropy of the full system, up to some small corrections. The second term becomes negligible compared to the first in the thermodynamic limit of particle number $N \rightarrow \infty.$ The last term represents a finite-size effect, which for small number of bins $m$ (or equivalently, large bin sizes) is also small in comparison with the first term. We will see in the next section that these corrections are an artifact of taking the determinate initial conditions: they disappear in case of indeterminate initial conditions.

\subsection{Dynamics of $S_{xE}$ for indeterminate initial conditions}\label{sec:starting_in_microcanonical}

Let us define the phase-space density of a microcanonical ensemble as
\[
\rho^{(\mathrm{micro})}_{E,\Delta E}(\bx,\bp)\equiv\begin{cases}
      \frac{\mu}{V_{E,\Delta E}}, & (\bx,\bp)\in P_{E,\Delta E} \\
      0, & (\bx,\bp)\not\in P_{E,\Delta E}
   \end{cases}.
\]
Equivalently, we can write
\[
\rho^{(\mathrm{micro})}_{E,\Delta E}=\frac{\mu\,\hat{P}_{E,\Delta E}} {V_{E,\Delta E}}.
\]
It is easy to see that for this microcanonical ensemble, as long as $dE\ll\Delta E$ (i.e., the energy coarse-graining is fine enough to determine the width of the ensemble), $S_{xE}$ gives the microcanonical entropy:
\[\label{eq:microcanonical_entropy}
S_{xE}\Big(\rho^{(\mathrm{micro})}_{E,\Delta E}\Big)=\ln V_{E,\Delta E}= S_{\mathrm{micro}}(E),
\]
where $\rho(E)$ denotes the energy density of states.

More generally, in Appendix~\ref{app:proofs} we prove a theorem that gives meaning to all stationary states: For phase-space densities that are mixtures of energy macrostates (such as the  microcanonical and canonical ensembles), $S_{xE}$ gives the thermodynamic entropy:
\begin{theorem}\label{thm:stationary_states} For phase-space density of form
$\rho=\mu\sum_E f(E)\hat{P}_E$, where $f(E)$ is any function of energy $E$ normalized as $\sum_E f(E)V_E =1$,
\[
S_{xE}(\rho)= S_{O(\C_H)} = S_{\mathrm{th}}(\rho).
\]
\end{theorem}

Now for the dynamics. We consider an initial state at some time $t<0$ that is a microcanonical state of $N$ particles contained in the left part of the box, and none in the other half, which we can denote as a microcanonical ensemble of the initial Hamiltonian, 
\[\label{eq:initial_state}
\rho_0\equiv\rho^{(\mathrm{micro};H_1)}_{E,\Delta E}.
\]
Since this is a stationary state of $H_1$, $\rho(t)=\rho_0$ for $t<0$.

Similar to Eq.~\eqref{eq:initialpoint}, we find that for $t<0$,
\[\label{eq:initial_micro}
S_{xE}(\rho_t)=\ln V_{E,\Delta E}^{(H_1)}\equiv S_{\mathrm{micro}}^{(H_1)}(E)=S_{\mathrm{micro}}(E,\tfrac{L}{2},N).
\]
(See Appendix~\ref{app:evolution} for details.) In other words, $S_{xE}$ of the initial state is equal to the microcanonical entropy of the first half of the box.

At time $t=0$, the available phase-space suddenly expands, and phase-space density $\rho(t)$ starts to explore the full extent of it. That first leads to a quick increase of entropy for the same reasons as for the case of determinate conditions, i.e., because the phase-space denstity will wander into the largest macrostate. However, due to mixing~\cite{arnold1968ergodic}, with time the positions of the particles become increasingly uncertain, and the phase-space density becomes uniformly smeared over all points in each energy shell. This is depicted as stages $2 \rightarrow 3$ in Fig.~\ref{Fig:phasespaceevolution}.

Mathematically, this means that the phase-space density converges to microcanonical state of the second Hamiltonian,
\[\label{eq:limit_nonexact}
\lim_{t\rightarrow +\infty}\rho(t)=\rho^{(\mathrm{micro};H_2)}_{E,\Delta E}.
\]

According to Theorem~\ref{thm:stationary_states}, $S_{xE}$ of this state must be equal to the microcanonical entropy of the entire box,\footnote{In order for this convergence to hold, we must take the width of energy coarse graining to be small, but non-zero, $0<dE\ll \Delta E$. $0<dE$ comes from the fact that Eq.~\eqref{eq:limit_nonexact} does not hold in a strict mathematical sense (there is however a way to write an exact mathematical statement, on which we will not elaborate here~\cite{arnold1968ergodic,volovich2009time}). This is because the phase-space density behaves like an incompressible fluid and as such it never uniformly fills up the energy shells of phase-space. However, it becomes dense in each shell, meaning that after waiting a long time, from the coarse-grained description given by non-zero $dE$ there will be no observable difference between the real microcanonical state and a state that is dense in an energy shell. And this is enough for the entropy $S_{xE}$ to not to register a difference between such states. We are taking $dE\ll \Delta E$ because we want Eq.~\eqref{eq:microcanonical_entropy} to hold.}
\[\label{eq:limitt}
\lim_{t\rightarrow +\infty}S_{xE}(t)=\ln V_{E,\Delta E}^{(H_2)}\equiv S_{\mathrm{micro}}^{(H_2)}(E)=S_{\mathrm{micro}}(E,L,N).
\]

Clearly, one can generalize this to any initial ensembles, since due to mixing, any initial phase-space density will become a stationary state of Hamiltonian $H_2$, in a sense of Theorem~\ref{thm:stationary_states}.

Having indeterminate initial conditions therefore results in $S_{xE}$ converging to thermodynamic entropy exactly, without the corrections of Eq.~\eqref{eq:limittpoint}.

\subsection{Simulations}

To support our analytical arguments, we have performed a simulation of a thermodynamic system of gas in $d=2$ spatial dimension, and for 4 and 16 partitions. The case of 4 partitions is depicted in Fig.~\ref{Fig:evolution}. 

We take $N=64$ particles of identical mass $m=1$, initialize them in the lower left corner of size $8\times 8$, within the full box of size $16\times 16$ with periodic boundary conditions. The velocity of each particle has been randomly drawn from the Normal distribution. Particles interact via a Lennard-Jones potential
\[
u(r)=\varepsilon\ \bigg[\Big(\frac{r_m}{r}\Big)^{12}-2\Big(\frac{r_m}{r}\Big)^{6}\bigg],
\]
where we took $\varepsilon=1/120$ and $r_m=1$ as parameters of the model, and $r$ denotes distance between each two particles. Particles are then evolved via a velocity Verlet algorithm with time step $10^{-4}$.

As particles and heat spread from one region to the others, entropy $S_{xE}$ grows from thermodynamic entropy of the first bin to the thermodynamic entropy of the full system, up to some finite-size corrections as expected from Eq.~\eqref{eq:limittpoint}, effectively modelling thermalization of an expanding gas.

\subsection{Interpretation}\label{sec:interpretationofSxE}

We have shown that $S_{xE}$, which is well-defined out of equilibrium, corresponds to the thermodynamical entropy of the initial region if all the particles are contained within this region, and it grows to thermodynamic entropy of the full system, for both cases of determinate and indeterminate initial conditions.

Physically, we can interpret $S_{xE}$ as a measure of equilibrium between the different regions defined by the coarse-graining in local particle numbers, but not necessarily as a measure of thermal equilibrium. If particles are uniformly distributed between the regions (the number of particles in each region being proportional the size of the region), $S_{xE}$ is equal to thermodynamic entropy. When $S_{xE}$ is low, it means that many particles are contained in one or a few small regions, and the system is therefore in a highly non-equilibrium state. $S_{xE}$ is therefore a measure of uniformity of particle density.

To illustrate this mathematically, for a single point in phase-space $(\bx,\bp)(t)\in P_{(n_1,\cdots,n_m)E}$, $S_{xE}$ gives value
\[
S_{xE}(t)=\ln V_{(n_1,\cdots,n_m)E},
\]
which is exactly the thermodynamic entropy that one would assign to a system of $N$ particles distributed into $m$ bins, in the situation where the energy is allowed to be exchanged between the bins, but particle number in each bin is fixed. 

To imagine this quantity operationally, imagine a situation where an observer/experimenter inserts elastic membranes in between the bins at time $t$ (infinitely quickly), which allow for energy transfer but not for particle transfer, and then waits until the system relaxes. The value of thermodynamic entropy he or she would assign to that system after it relaxes to thermal equilibrium is $S_{xE}(t)$.

This brings out problematic feature of this entropy, which is that the coarse-graining in global energy does not distinguish any microstates that differ in energy locally. This means that the same entropy will be associated to any microstate that has $E_1+E_2=E$, with $E_1$ and $E_2$ being the energy of the left region and of the right region respectively. As an example, imagine a situation with $10$ particles in the left region, and $10$ particles in the right region. Situations when the left part is really hot and the right part really cold, and when both parts have equal temperatures, will both have the same associated $S_{xE}=\ln V_{(10,10)E}$ as long as the total energy $E$ is the same. This also means that $S_{xE}$ is not in general additive. \footnote{The studied example of an expanding gas was a special case, because there were no particles in the right part of the box, therefore this part did not contribute to the total entropy (explained in our operational view: no energy could be transferred if there were no particles on the right when the elastic membrane was inserted).}

As we'll see, factorized Observational entropy $S_F$ solves this problem, by building on the definition of $S_{xE}$ but using the {\em local} energy coarse-grainings, instead of the global energy coarse-graining.

\section{$S_F$: entropy of measuring local particle numbers and local energies}\label{sec:SF}

In the same 1-D system introduced in the last section, we define $\text{local~particle~number-local~energy}$ macrostates as
\[\label{eq:localnlocalE}
\begin{split}
&P_{\bn\bE}\equiv \{(\bx,\bp)|n_1 \text{ particles with position } x \in [L_1,L_1\!+\!\Delta x),\\
&\ \dots, n_m \text{ particles with position } x \in [L_1\!+\!(m\!-\!1)\Delta x,L_2),\\
& \text{energy of the bin } [L_1,L_1\!+\!\Delta x) \text{ is between } [E_1,E_1+dE),\\
&\dots,\text{energy of the bin } [L_1\!+\!(m\!-\!1)\Delta x,L_2)\text{ is between }\\ & [E_m,E_m+dE)\},
\end{split}
\]
where $\bn=(n_1,\dots,n_m)$ and $\bE=(E_1,\dots,E_m)$ are vectors representing the local particle numbers and local energies. $N=n_1+\cdots+n_m$, $E=E_1+\cdots+E_m$, and $L=L_2-L_1$ are the total particle number, the total energy, and the total spatial volume (length of the box). This can be generalized to any spatial dimension $d$ and spatial volume ${\cal V}$, with number of bins defined as $m=\frac{{\cal V}}{\Delta x^d}$.

Observational entropy with coarse-graining $\C_{X\bE}=\{P_{\bn\bE}\}_{\bn\bE}$ corresponding to measuring local particle numbers and local energies
\[\label{eq:Sxe_definition}
S_{F}(t)\equiv S_{O(\C_{X\bE})}(t)=-\sum_{\bn, \bE}p_{\bn \bE}(t)\ln\frac{p_{\bn \bE}(t)}{V_{\bn \bE}}
\]
satisfies all properties required of \emph{dynamical thermodynamic entropy}.

Assuming indistinguishable particles of a sufficiently dilute and weakly interacting gas, volume of macrostate $P_{\bn\bE}$ is equal to the product of local volumes, $V_{\bn\bE}=V_{n_1E_1}\cdots V_{n_mE_m}$ (see Appendix~\ref{app:propertiessf}). As a result, for a single point in phase-space $(\bx,\bp)(t)\in P_{\bn\bE}$,
\[
S_{F}(t)=\ln V_{\bn\bE}=\sum_{k=1}^m S_{\mathrm{micro}}(E_k,\Delta x^d,n_k),
\]
where $S_{\mathrm{micro}}(E_k,\Delta x^d,n_k)=\ln V_{n_kE_k}$ is the microcanonical entropy of the $k^{\mathrm{th}}$ bin. In other words, unlike $S_{xE}$, for a single point in phase-space $S_F$ is equal to the sum of local microcanonical entropies, and is therefore also additive on independent systems.

In the long time limit, this entropy converges to the total microcanonical entropy,
\[\label{eq:limittpointfoe}
\begin{split}
\lim_{t\rightarrow +\infty}S_{F}(t)&=S_{\mathrm{micro}}(E,{\cal V},N)+\ln (2\pi N) -m\ln\frac{2\pi N}{m}\\
&-\frac{m-1}{2}\ln\frac{d}{2},
\end{split}
\]
up to some corrections which are about a factor of two larger than those for $S_{xE}$. (For $d=1$, ${\cal V}=L$.) These corrections are due to determinate initial conditions (starting as a single point in phase-space), and they disappear in the long time limit when starting with indeterminate conditions due to mixing. The proof is very similar to the proof of the same property for $S_{xE}$, Eqs.~\eqref{eq:limittpoint} and \eqref{eq:limitt}, and can be also found in Appendix~\ref{app:propertiessf}.

Similar to $S_{xE}$, $S_F$ is always defined even for systems out of equilibrium, and for systems with indeterminate initial conditions, given by a phase-space density $\rho$.

Since the coarse-graining used by $S_F$ is finer than that of $S_{xE}$, from Theorem~\ref{thm:monotonic} follows that
\[
S_F(t)\leq S_{xE}(t).
\]
This holds up to some finite size corrections, which comes from the fact that $S_F$ uses local Hamiltonians\footnote{For the definition of local Hamiltonians, see Appendix~\ref{app:propertiessf}.} that ignore interaction between the bins.

$S_F$ is a measure of thermal equilibrium between the regions. If the value of $S_F$ is high, it means that both particles and energy are uniformly distributed across the entire system. If the value of $S_F$ is low, it means that there are regions with a much higher density of particles and energy compared to other regions, which corresponds to a highly non-equilibrium state. $S_F$ is therefore a measure of uniformity of both particle and energy density.

Operationally, $S_F(t)$ corresponds to a situation when an observer/experimenter inserts walls in between the bins at time $t$ (infinitely quickly), which do not allow for either energy or particle transfer between the bins. Waiting until the system relaxes means that each region is in a thermal equilibrium with itself, but not with other regions. The thermodynamic entropy the experimenter would assign to that system after this type of relaxation is $S_{F}(t)$.

\begin{table*}[htbp]
\centering
\caption{\label{tab:quantumclassrelations}  Classical and quantum descriptions of an isolated physical system.}
\begin{ruledtabular}
\begin{tabular}{ll}
\rowcolor{red!10} {\bf Classical} &  {\bf Quantum} \\
\hline
\rowcolor{blue!10} {\bf phase-space} $\Gamma$  & {\bf Hilbert space} $\HS$\\
~~~defines a classical system, &  ~~~defines a quantum system,  \\
~~~is a space of all possible classical states, &  ~~~is a space of all possible quantum states,  \\
~~~all states are orthogonal, &  ~~~employs Hilbert space inner product,\\
~~~employs $L^2$-inner product for observables &  ~~~employs Hilbert-Schmidt inner product for observables.\\
\hline
\rowcolor{blue!10} {\bf Point in phase-space} $(\bx,\bp)$  & {\bf Vector in Hilbert space} $\ket{\psi}$ -- wave-function \\
~~~describes a state of a classical system, &  ~~~describes a state of a quantum system,  \\
~~~is evolved through Hamilton's equations of motion, &  ~~~is evolved through Schr\"odinger equation,  \\
~~~is also called a microstate. &  ~~~is also called a microstate.  \\
\hline
\rowcolor{blue!10} {\bf phase-space density} $\rho$  & {\bf Density matrix} $\R$ \\
~~~describes a classical system of indeterminate state, &  ~~~describes a quantum system of indeterminate state, \\
~~~is evolved through Liouville's equation. &  ~~~is evolved through von Neumann equation.  \\
~~~Point in phase-space is described by a $\delta$-function, &  ~~~Vector in Hilbert space is described by a rank-1 \\
~~~$\rho(\tilde\bx,\tilde\bp)=\delta(\tilde\bx-\bx,\tilde\bp-\bp)$. &  ~~~density matrix, $\R^2=\R$.  \\
\hline
\rowcolor{blue!10} {\bf Coarse-graining} $\C=\{P_i\}_i$  & {\bf Coarse-graining} $\C=\{\HS_i\}_i$ \\
~~~is a partition of phase-space, $\Gamma=\bigcup_i P_i$. &  ~~~is a partition of Hilbert space, $\HS=\bigoplus_i \HS_i$. \\
Equivalently,  $\C=\{\P_i\}_i$ &  Equivalently,  $\C=\{\P_i\}_i$ \\
~~~is a complete set of orthogonal window functions, $\sum_i\P_i=1$. &  ~~~is a complete set of orthogonal projectors, $\sum_i\P_i=\hat{I}$.  \\
~~~$P_i$ or equivalently $\P_i$ is called a macrostate. &  ~~~$\HS_i$ or equivalently $\P_i$ is called a macrostate.  \\
~~~A joint coarse-graining of multiple coarse-grainings always & ~~~A joint coarse-graining exists only if the coarse-grainings  \\
~~~exists. &  ~~~commute.  \\
\hline
\rowcolor{blue!10} {\bf Observable} $A$  & {\bf Observable} $\hat{A}$ \\
~~~is a real-valued function $A:(\bx,\bp)\rightarrow a$ acting on  &  ~~~is a Hermitian operator acting on a Hilbert space, \\

~~~a phase-space, &  ~~~ \\
~~~admitting spectral decomposition $A=\sum_a a \P_a$,  &  ~~~admitting spectral decomposition $\hat{A}=\sum_a a \P_a$, \\
~~~with expectation value $\mean{A}=(A,\rho)$, &  ~~~with expectation value $\mean{\hat{A}}=\tr[\hat{A}\R]$,  \\
~~~where the probability of observing $a$ is $p_a=(\P_a,\rho)$, &  ~~~where the probability of observing $a$ is $p_a=\tr[\P_a\R]$,  \\
~~~and coarse-graining given by an observable is $\C_A=\{\P_a\}_a$. &  ~~~and coarse-graining given by an observable is $\C_{\hat{A}}=\{\P_a\}_a$.\\
\hline
\rowcolor{blue!10} {\bf Observational entropy} $S_{O(\C_1,\dots,\C_n)}(\rho)=-\sum_\bi p_\bi\ln \frac{p_\bi}{V_\bi}$,  & {\bf Observational entropy} $S_{O(\C_1,\dots,\C_n)}(\R)=-\sum_\bi p_\bi\ln \frac{p_\bi}{V_\bi},$ \\
~~~where $p_{\bi}=\int_{\Gamma}\rho \P_{i_1}\cdots\P_{i_n} d\bx d\bp$ is a probability of belonging &  ~~~where $p_{\bi}=\tr[\P_{i_n}\cdots\P_{i_1}\R \P_{i_1}\cdots\P_{i_n}]$ is a probability \\
~~~to multi-macrostate $\bi=(i_1,\dots,i_n)$, &  ~~~of belonging to multi-macrostate $\bi=(i_1,\dots,i_n)$, \\
~~~and  $V_{\bi}=\mu\int_{\Gamma} \P_{i_1}\cdots\P_{i_n} d\bx d\bp$ is volume of multi-macrostate $\bi$. &  ~~~and  $V_{\bi}=\tr[\P_{i_n}\cdots\P_{i_1}\cdots\P_{i_n}]$ is volume of multi-macrostate $\bi$.  \\
~~~A point in phase-space belongs only into a single &  ~~~A wave-function can span over several macrostates. \\
~~~macrostate. &  \\
~~~$S_{O(\C_1,\C_2)}=S_{O(\C_2,\C_1)}$ always holds. &  ~~~$S_{O(\C_1,\C_2)}= S_{O(\C_2,\C_1)}$ holds for commuting coarse-grainings, \\
~~~ &  ~~~but does not hold for non-commuting coarse-grainings.\\
~~~$S_G(\rho)\leq S_{O(\C_1,\dots,\C_n)}(\rho)\leq \ln V$ &  ~~~$S_{V\!N}(\R)\leq S_{O(\C_1,\dots,\C_n)}(\R)\leq \ln \dim \HS$ \\
~~~$S_{O(\C_1,\dots,\C_n)}(\rho)\leq S_{O(\C_1,\dots,\C_{n-1})}(\rho)$ &  ~~~$S_{O(\C_1,\dots,\C_n)}(\R)\leq S_{O(\C_1,\dots,\C_{n-1})}(\R)$ \\
\hline
\rowcolor{blue!10} {\bf Dynamical thermodynamic entropy} & {\bf Dynamical thermodynamic entropy} \\
\rowcolor{blue!10} $\boldsymbol{S}_{\boldsymbol{x}\boldsymbol{E}}\equiv S_{O(\C_X,\C_H)}=-\sum_{\bn, E}p_{\bn E}\ln\frac{p_{\bn E}}{V_{\bn E}}$,  & $\boldsymbol{S}_{\boldsymbol{x}\boldsymbol{E}}\equiv S_{O(\C_{\hat{X}},\C_{\hat{H}})}=-\sum_{\bn, E}p_{\bn E}\ln\frac{p_{\bn E}}{V_{\bn E}}$, \\
~~~where $\C_X=\{P_{\bn}\}_{\bn}$ is a set of a local particle number &  ~~~where $\C_{\hat{X}}=\{\P_{\bn}\}_{\bn}$ is a set of a local particle number \\
~~~macrostates, and $\C_{H}=\{P_{E}\}_{E}$ is a set of energy macrostates, &  ~~~macrostates, and $\C_{\hat{H}}=\{\P_{E}\}_{E}$ is a set of energy macrostates, \\
~~~which are defined as energy shells of the Hamiltonian, &  ~~~which are defined as projectors onto energy eigenstates, \\
~~~$P_{E}= \{(\bx,\bp)|E\!\leq\! H(\bx,\bp)\!<\!E+dE\}$ ($dE$ independent of $E$). &  ~~~$\P_E=\pro{E}{E}$, $\hat{H}\ket{E}=E\ket{E}$.\\
~~~Converges to thermodynamic entropy. Is not additive. &  ~~~Converges to thermodynamic entropy. Is additive.\\
~~~ & ~~~Is not additive if a finite resolution in energy $\Delta E$ is used.\\
& \cellcolor{blue!10} $\boldsymbol{S}_{\boldsymbol{E}\boldsymbol{x}}\equiv S_{O(\C_{\hat{H}^{(\Delta E)}},\C_{\hat{X}})}=-\sum_{\bn, E}p_{E\bn}\ln\frac{p_{E \bn}}{V_{E \bn}}$, \\
&  ~~~where $\C_{\hat{H}^{(\Delta E)}}=\{\P_{E,\Delta E}\}_{E}$ is a set of energy macrostates, \\
&  ~~~which are defined as projectors onto coarse-grained energy  \\
&  ~~~macrostates, $\P_{E,\Delta E}=\sum_{E\leq \tilde{E}<E+\Delta E}\pro{\tilde{E}}{\tilde{E}}$.\\
\rowcolor{blue!10} {\bf Dynamical thermodynamic entropy} & {\bf Dynamical thermodynamic entropy} \\
\rowcolor{blue!10} $\boldsymbol{S}_{F}\equiv S_{O(\C_{X\bE})}=-\sum_{\bn, E}p_{\bn \bE}\ln\frac{p_{\bn \bE}}{V_{\bn \bE}}$,  & $\boldsymbol{S}_{F}\equiv S_{O(\C_{\hat{H}_1}\otimes\cdots\otimes \C_{\hat{H}_m})}=-\sum_{\bE}p_{\bE}\ln\frac{p_{\bE}}{V_{\bE}}$, \\
~~~where $\C_{X\bE}=\{P_{\bn\bE}\}_{\bn\bE}$ is a set of a local particle number &  ~~~where $\C_{\hat{H}_1}\otimes\cdots\otimes \C_{\hat{H}_m}$ is a set of local energy\\
~~~and local energy macrostates.&  ~~~macrostates. \\
~~~Converges to thermodynamic entropy. Is additive.&  ~~~Converges to thermodynamic entropy. Is additive.\\
~~~&  ~~~Is modified to
include coarse-graining in local particle
\\
~~~&  ~~~numbers if a finite resolution in energy $\Delta E$ is used.$^{\ref{foot:deltalarger}}$
\\
\end{tabular}
\end{ruledtabular}
\end{table*}

\section{Comparison of classical and quantum Observational entropy}\label{sec:class_quant}

Having investigated in detail particular coarse-grainings that connect classical Observational entropy to thermodynamics, we compare Observational entropy, and these particular coarse-grainings, across the classical and quantum case.

In the same way that a point in phase-space describes a classical system, a vector in a Hilbert space describes a quantum system. This state-vector (or wave-function) encodes every property of a quantum system and is evolved using the Schr\"odinger equation. More generally, a quantum system with indeterminate initial condition is described by a density matrix $\R$, which is a positive semi-definite operator acting on the Hilbert space. The density matrix is a quantum equivalent of phase-space density, and is evolved through the von Neumann equation, as compared to Liouville's equation in classical system.

In classical systems, the coarse-graining is defined as a partitioning of phase-space. These partitions/regions --- subsets of phase-space --- are called the macrostates. To define coarse-graining in quantum physics, we have to partition a Hilbert space. However, partitions of Hilbert space are not composed of subsets but of sub{\em spaces} combined in a direct sum:  $\HS=\HS_1\oplus\HS_2\oplus\cdots$. We can form these subspaces for example by choosing a basis of the Hilbert space, and putting every basis vector into a group. A subspace is then created by taking all the linear combinations of vectors in the group. For example, real three-dimensional space could be partitioned as $\mathbb{R}^3=\mathbb{R}^2\oplus \mathbb{R}$, where subspace $\mathbb{R}^2=\mathrm{span}\{(1,0,0),(0,1,0)\}$ describes the {x-y} plane, and $\mathbb{R}=\mathrm{span}\{(0,0,1)\}$ the {z-axis}. We call these subspaces macrostates, and their collection a coarse-graining, $\C=\{\HS_i\}_i$.

As we did with the classical system in Eq.~\eqref{eq:window_func}, also here we can switch from describing the coarse-graining as a collection of subspaces, and describe it as a collection of projectors that project onto these subspaces instead. In other words, for each subspace $\HS_i$ there exists a unique projector $\P_i$ that projects onto this subspace, and the coarse-graining $\C=\{\P_i\}_i$ is then defined as a complete set of orthogonal projectors ($\P_i^2=\P_i$, $\P^\dag=\P$, $\P_i\P_j=\delta_{ij}\P_i$, $\sum_i\P_i=\hat{I}$).

With an ordered set of coarse-grainings $(\C_1,\dots,\C_n)$, we define volume of a multi-macrostate $\bi=(i_1,\dots,i_n)$ as
\[
V_\bi=\tr[\P_{i_n}\cdots\P_{i_1}\cdots\P_{i_n}],
\]
and the probability of being in a macrostate as
\[
p_\bi(\R)=\tr[\P_{i_n}\cdots\P_{i_1}\R\P_{i_1}\cdots\P_{i_n}].
\]
This probability can be also interpreted as a probability of obtaining the sequence of outcomes $\bi$ when performing a sequence of measurements in measurement bases $(\C_1,\dots,\C_n)$ on the system.

The (quantum) Observational entropy is defined identically to the classical case, Eq.~\eqref{eq:definition_of_generalOE}, but with the use the above definitions for $V_\bi$ and $p_\bi(\R)$:
\[
S_{O(\C_1,\dots,\C_n)}(\R)=-\sum_\bi p_\bi(\R)\ln \frac{p_\bi(\R)}{V_\bi}.
\]

We compare descriptions of classical and quantum systems in Table~\ref{tab:quantumclassrelations}. Notably, coarse-grainings in the classical scenario always commute, while in quantum scenario they need not. As a consequence, a joint coarse-graining may not exist in the quantum case, and switching order of non-commuting coarse-grainings leads to different Observational entropies. Surprisingly, this does not affect many other properties, as theorems equivalent to \ref{thm:Boltzman_generalization}-\ref{thm:stationary_states} still hold. 

More stark differences appear when defining thermodynamically relevant coarse-grainings. Due to the non-commutativity of coarse-grainings, there are two quantum versions of $S_{xE}$, one that corresponds to first measuring the the local particle numbers, and then energy (denoted $S_{xE}$), and one that corresponds to first measuring energy and then local particle numbers (denoted $S_{Ex}$). The behavior of quantum $S_{xE}$ or $S_{Ex}$ further depends heavily on whether the energy resolution is finite or infinite, i.e whether it is assumed that the experimenter cannot distinguish between two energy eigenstates with energies $|E_1-E_2|<\Delta E$. Since classical energy eigenstates do not exist, the classical $S_{xE}$ inherently entails such finite resolution in its definition. 

This leads to a complex and rather subtle set of correspondences.  When $\Delta E$ is non-negligible (relative say to the temperature of the system) and the temperature is high, these differences are negligible, and classical and quantum $S_{xE}$ and $S_{Ex}$ all behave in the same way. ({\em However} it should be noted that in this case -- as touched upon in Sec.~\ref{sec:interpretationofSxE} and detailed in Appendix~\ref{app:additivityofsxe} -- $S_{xE}$ is not always additive.) For $\Delta E > 0$ at {\em low} temperatures and/or {\em small} bin sizes $\Delta x$, $S_{xE}$ and $S_{Ex}$ can differ appreciably due to large effects of non-commutativity. Finally, for $\Delta E=0$ (requiring the quantum case), $S_{xE}$ and $S_{Ex}$ are quite distinct: $S_{xE}$ behaves much like quantum $S_F$ (and is additive; see Appendix~\ref{app:additivityofsxe}) while $S_{Ex}$ is constant in time, equal to thermodynamic entropy $S_{E}$, and zero on any global energy eigenstate.

Turning to $S_F$, there is no issue of commutation, but the quantum case again gives an option of $\Delta E=0$ or finite $\Delta E.$ The $\Delta E=0$ case defined $S_F$ in~\cite{safranek2019letter,safranek2019long}, and was shown to be very similar to quantum $S_{xE}$. One could also define a quantum $S_F$ with $\Delta E> 0$ to closely correspond to the classical case.\footnote{\label{foot:deltalarger}For $\Delta E>0$ it is necessary to include also coarse-graining in local particle numbers, in order for this entropy to give the correct thermodynamic values. The definition then reads $\boldsymbol{S}_{F}\equiv S_{O(\C_{\hat{N}_1}\otimes\cdots\otimes \C_{\hat{N}_m},\C_{\hat{H}_1^{(\Delta E)}}\otimes\cdots\otimes \C_{\hat{H}_m^{(\Delta E)}})} =-\sum_{\bn,\bE}p_{\bn\bE}\ln\frac{p_{\bn\bE}}{V_{\bn\bE}}$. For $\Delta E=0$ this was not necessary, because measuring local energy eigenstates uniquely determined the local particle numbers, for particle conserving Hamiltonians.}  In all cases $S_F$ is additive.

Generally, one could expect differences between classical and quantum definitions of entropies for systems that where value of Planck's constant is comparable to quantities with the same dimensions which appear in the problem~\cite{balian2007microphysics324}.

\section{Conclusion and outlook}
This paper has discussed in detail the framework of classical ``Observational entropy," an idea previously defined in the quantum context~\cite{safranek2019letter,safranek2019long}. This quantity is defined precisely in or out of equilibrium, is generically non-decreasing, and corresponds to thermodynamic entropy in equilibrium.
Our treatment has aimed to define this quantity clearly and rigorously, while also exhibiting in detail three core sets of relations.

First, the treatment shows how Observation entropy generalizes and interpolates between classical ``Gibbs'' and ``Boltzmann'' entropies.  The latter is often thought of as Gibbs entropy where equal probability is attributed to all microstates compatible with a given set of macroscopic constraints.  Here, we see Boltzmann entropy as a limit of Observational entropy in which all probability is attributed to a single microstate, while Gibbs entropy appears in the limit in which the coarse-graining is as fine as possible so that each microstate constitutes a macrostate.

Second, Observational entropy has both an information-theoretic and a thermodynamic interpretation.  Every additional measurement on a system -- corresponding to an additional coarse-graining -- that better pins down a particular subset of phase-space yields more information in a clear and quantifiable way. At the same time, here (Eq.~\eqref{eq:limittpoint}) and in previous work~\cite{safranek2019letter,safranek2019long} we have shown that with an appropriate choice of coarse-graining, Observational entropy describes dynamically thermalization of a gas, converging to thermodynamic entropy when the system equilibrates. These coarse-graining also have an elegant interpretation out of equilibrium, corresponding to the thermodynamic entropy the system would attain in equilibrium {\em if} particular constraints were imposed on the sharing of particles and energy across spatial regions.

Third, this work demonstrates a clear and well-defined correspondence between the quantum and classical cases. By defining projection operators that on Hilbert space and classical phase-space, respectively, carefully defining volume units for classical phase-space, and suitably defining quantum and classical density operators, we show that formulas and most theorems concerning Observational entropy carry over directly between the classical and quantum case.  The only key fundamental differences arise in the non-commutation of coarse-grainings in the quantum case, and from the non-existence of a direct classical analogue of quantum energy eigenstates. In contrast, quantum entanglement has no direct classical analog, so entanglement entropy lacks -- to our knowledge -- a clear correspondence with a classical quantity that behaves similarly. 

Thus we view Observational entropy as a framework that relates and unifies a number of disparate views and definitions of entropy, as well as provides useful new perspectives on the connection between quantum and classical treatments of phenomena.  We hope that this treatment proves both illuminating and useful in understanding how entropy manifests and operates in a wide range of physical systems.  

\acknowledgements{We are grateful to Dana Faiez and Joseph Schindler for helpful discussions. 
This research was supported by the Foundational Questions Institute (FQXi.org), of which AA is Associate Director, and by the Faggin Presidential Chair Fund.}

\appendix
\section{Proofs}\label{app:proofs}

We are going to use {\bf Jensen's inequality}:
Let $f$ be a strictly concave function, $0\leq a_i \leq 1$, $\sum_i a_i=1$. Then for any $b_i\in\mathbb{R}$,
\[
f\big(\sum_i a_i b_i\big)\geq \sum_i a_i f(b_i).
\]
$f(\sum_i a_i b_i)= \sum_i a_i f(b_i)$ if and only if
$
(\forall i,j|a_i\neq 0, a_j\neq 0)(b_i=b_j).
$

Now we proceed with proofs of the paper's theorems.
\begin{proof} (Theorem~\ref{thm:Boltzman_generalization}) (Observational entropy is a generalization of the Boltzmann entropy)
Clearly, for a single point in phase-space $(\bx,\bp)\in P_i$, we have $p_i=1$ and $p_j=0$ for $j\neq i$. Therefore
\[
S_{O(\C)}(\rho)= \ln V_i=S_B(\rho).
\]
\end{proof}

\begin{proof} (Theorem~\ref{thm:monotonic}) (Observational entropy is a monotonic function of the coarse-graining.)
$\C_1\hookrightarrow \C_2$ means that $P_{i_1}=\bigcup_{i_2\in I^{(i_1)}}P_{i_2}$ for each $i_1$, therefore $p_{i_1}=\sum_{i_2\in I^{(i_1)}}p_{i_2}$ and $V_{i_1}=\sum_{i_2\in I^{(i_1)}}V_{i_2}$. Then from Jensen's inequality applied on concave function $f(x)=-x\ln x$, taking $a_{i_2}=\frac{V_{i_2}}{V_{i_1}}$ and $b_{i_2}=\frac{p_{i_2}}{V_{i_2}}$, we have
\[
\begin{split}
 S_{O(\C_1)}(\rho)&=-\sum_{i_1}p_{i_1}\ln\frac{p_{i_1}}{V_{i_1}}\\
&=-\sum_{i_1}\sum_{i_2\in I^{(i_1)}}p_{i_2}\ln\frac{\sum_{i_2\in I^{(i_1)}}p_{i_2}}{V_{i_1}}\\
&=\sum_{i_1}V_{i_1}\bigg(-\sum_{i_2\in I^{(i_1)}}\frac{V_{i_2}}{V_{i_1}}\frac{p_{i_2}}{V_{i_2}}\ln\sum_{i_2\in I^{(i_1)}}\frac{V_{i_2}}{V_{i_1}}\frac{p_{i_2}}{V_{i_2}}\bigg)\\
&\geq \sum_{i_1}V_{i_1}\bigg(-\sum_{i_2\in I^{(i_1)}}\frac{V_{i_2}}{V_{i_1}}\frac{p_{i_2}}{V_{i_2}}\ln\frac{p_{i_2}}{V_{i_2}}\bigg)\\
&=-\sum_{i_1}\sum_{i_2\in I^{(i_1)}}p_{i_2}\ln\frac{p_{i_2}}{V_{i_2}}=S_{O(\C_2)}(\rho).
\end{split}
\]

The equality conditions from the Jensen's inequality show that $S_{O(\C_1)}(\R)=S_{O(\C_2)}(\R)$ if and only if
\[\label{eq:equality_monotonic}
\begin{split}
&(\forall i_1|V_{i_1}\neq 0)(\forall i_2,\tilde{i}_2\in I^{(i_1)}|V_{i_2}\neq 0, V_{\tilde{i}_2}\neq 0 )\\
&\left(\frac{p_{i_2}}{V_{i_2}}=\frac{p_{\tilde{i}_2}}{V_{\tilde{i}_2}}=c^{(i_1)}\right). 
\end{split}
\]
To determine the constant $c^{(i_1)}$ we multiply the equation by $V_{{i}_2}$ and sum over all $\forall i_2\in I^{(i_1)}$, which gives
\[
c^{(i_1)}=\frac{p_{i_1}}{V_{i_1}}.
\]
And considering that for all $V_{i_2}=0$ also $p_{i_2}=0$, we can simplify Eq.~\eqref{eq:equality_monotonic}, and obtain that $S_{O(\C_1)}(\R)=S_{O(\C_2)}(\R)$ if and only if
\[
(\forall i_1|V_{i_1}\neq 0)(\forall i_2\in I^{(i_1)})\left(p_{i_2}=\frac{V_{i_2}}{V_{i_1}}p_{i_1}\right).
\]
\end{proof}

\begin{proof} (Theorem~\ref{thm:bounded_multiple}) (Observational entropy with multiple coarse-grainings is bounded)
Since Observational entropy with multiple coarse-grainings can be rewritten as Observational entropy with a joint coarse-graining,
\[\label{eq:joint_coarse}
S_{O(\C_1,\dots,\C_n)}(\rho)=S_{O(\C_{1,\dots,n})}(\rho),
\]
it is enough to prove the inequalities just for a single coarse-graining $\C$,
\[
S_{G}(\rho)\leq S_{O(\C)}(\rho)\leq \ln V.
\]

For the second inequality we define coarse-graining with a single element -- the entire phase-space -- $\C_\Gamma=\{\Gamma\}$. Clearly, this coarse-graining is coarser than any other coarse-graining, therefore from Theorem~\ref{thm:monotonic} we have
\[
S_{O(\C)}(\rho)\leq S_{O(\C_\Gamma)}(\rho)=-p_\Gamma\ln\frac{p_\Gamma}{V_\Gamma}=-1\ln\frac{1}{V}=\ln V.
\]

The first inequality comes from the fact that one can choose a coarse-graining where macrostates are single points in phase-space. This coarse-graining is finer than any other coarse-graining, and one can easily derive that the Observational entropy is then equal to the Gibbs entropy. By the same argument we can therefore obtain the first inequality. However, this argument does not give us the equality conditions, for which we will have to make a more elaborate derivation as follows.

To prove the first inequality and the equality conditions, we define the spectral decomposition of the phase-space density in its eigenvector projectors  $\rho=\sum_{(\bx,\bp)}\rho(\bx,\bp)\P_{(\bx,\bp)}$ (meaning $\rho(\tilde{\bx},\tilde{\bp})=\sum_{(\bx,\bp)}\rho(\bx,\bp)\P_{(\bx,\bp)}(\tilde{\bx},\tilde{\bp})$, where eigenvalues $\rho(\bx,\bp)$ do not have to be different for different $(\bx,\bp)$. The eigenvector projectors project onto infinitesimal regions surrounding $(\bx,\bp)$ are defined as
\[
\P_{(\bx,\bp)} (\tilde{\bx},\tilde{\bp})=\begin{cases}
      1, & (\tilde{\bx},\tilde{\bp})\in[\bx,\bx+d\bx)\times[\bp,\bp+d\bp), \\
      0, & \mathrm{otherwise}.
   \end{cases}
\]
We also define spectral decomposition of the density matrix in a form where the eigenvector projectors associated with the same eigenvalue are now grouped together, 
\[\label{eq:uniquelambdadecomposition}
\rho=\sum_{\lambda}\lambda\P_{\lambda},
\]
where eigenvalues $\lambda$ are now different from each other. This decomposition is unique. It follows that for each $(\bx,\bp)$ there exists $\lambda$ such that $\rho(\bx,\bp)=\lambda$.

Now recall Eq.~\eqref{eq:V_and_p}: the probability of a state being in a macrostate $i$ of volume $V_i=(\P_{i},\mu)$ is $p_i=(\P_{i},\rho)$.

Defining
\[
a_{(\bx,\bp)}^{(i)}\equiv\frac{\mu(\P_i,\P_{(\bx,\bp)})}{V_i},
\]
for $V_i\neq 0$ and $a_{(\bx,\bp)}^{(i)}\equiv 0$ for $V_\bi= 0$, and then using the spectral decomposition of $\rho$ we have
\[\label{eq:p_over_V}
\frac{p_i}{V_i}=\frac{(\P_i,\sum_{(\bx,\bp)}\rho(\bx,\bp)\P_{(\bx,\bp)})}{V_i}
=\sum_{(\bx,\bp)}\frac{\rho(\bx,\bp)}{\mu} a_{(\bx,\bp)}^{(i)}. 
\]
Since $\sum_{(\bx,\bp)}\P_{(\bx,\bp)}=1$,
\[\label{eq:sum_ax}
\sum_{(\bx,\bp)} a_{(\bx,\bp)}^{(i)}=1,
\]
and since $\sum_i\P_i=1$, also
\[\label{sum_V_ax}
\sum_i V_i a_{(\bx,\bp)}^{(i)}=(\mu,\P_{(\bx,\bp)})=\mu d\bx d\bp.
\]

A series of equalities and inequalities follow:
\[\label{seriesofinequalitiesforgibbs}
\begin{split}
&S_{O(\C_1,\dots,\C_n)}(\R)=-\sum_{i}p_i\ln \frac{p_i}{V_{i}}\\
&=-\sum_{i}V_{i}\frac{p_i}{V_{i}}\ln \frac{p_i}{V_{i}}\\
&=\sum_{i}V_{i}\left(-\sum_{(\bx,\bp)}\frac{\rho(\bx,\bp)}{\mu} a_{(\bx,\bp)}^{(i)}\ln \sum_{(\bx,\bp)}\frac{\rho(\bx,\bp)}{\mu} a_{(\bx,\bp)}^{(i)}\right)\\
&\geq \sum_{i}V_{i} \left(-\sum_{(\bx,\bp)}\frac{\rho(\bx,\bp)}{\mu} a_{(\bx,\bp)}^{(i)}\ln \frac{\rho(\bx,\bp)}{\mu}\right)\\
&=-\sum_{(\bx,\bp)}\left( \sum_{i}V_{i} a_{(\bx,\bp)}^{(i)}\right) \frac{\rho(\bx,\bp)}{\mu}\ln \frac{\rho(\bx,\bp)}{\mu}\\
&=-\int_\Gamma\frac{\rho(\bx,\bp)}{\mu}\ln \frac{\rho(\bx,\bp)}{\mu}\ \mu d\bx d\bp\equiv S_G(\rho).
\end{split}
\]
We have used Eqs.~\eqref{eq:p_over_V} and~\eqref{sum_V_ax} for the equalities, and applied Jensen's Theorem on function $f(x)=-x\ln x$ to derive the inequality. We have chosen $a_{(\bx,\bp)}\equiv a_{(\bx,\bp)}^{(i)}$ and $b_{(\bx,\bp)}\equiv\frac{\rho(\bx,\bp)}{\mu}$ for the Theorem.

According to Jensen's Theorem, the inequality becomes equality if and only if
\[\label{eq:first_eq_condition_old}
\begin{split}
&(\forall i|V_{i}\!\neq\! 0)(\forall (\bx,\bp),(\tilde{\bx},\tilde{\bp})| (\P_i,\P_{(\bx,\bp)})\!\neq\! 0,(\P_i,\P_{(\tilde{\bx},\tilde{\bp})})\!\neq\! 0)\\
&(\rho(\bx,\bp)=\rho(\tilde{\bx},\tilde{\bp})).
\end{split}
\]
That is, the inequality becomes equality when for a given index $i$, all eigenvector projectors $\P_{(\tilde{\bx},\tilde{\bp})}$ of the phase-space density such that $(\P_i,\P_{(\bx,\bp)})\neq 0$ have the same associated eigenvalue $\rho(\bx,\bp)$ with them. In other words, we can associate this unique eigenvalue to the index $i$ itself, $\rho_i \equiv\rho(\bx,\bp)$, where $\rho(\bx,\bp)$ is given by any representative $(\bx,\bp)$ such that $(\P_i,\P_{(\bx,\bp)})\neq 0$. Realizing that $(\P_i,\P_{(\bx,\bp)})\neq 0$ is the same as saying $P_i\cap P_{(\bx,\bp)}\neq \O$, or $\P_{(\bx,\bp)}\P_i\neq 0$ we can rewrite Eq.~\eqref{eq:first_eq_condition_old} as
\[\label{eq:conditionrewritten}
\begin{split}
&(\forall i|V_{i}\!\neq\! 0)(\forall (\bx,\bp),(\tilde{\bx},\tilde{\bp})| \P_{(\bx,\bp)}\P_i\neq 0,\P_{(\tilde{\bx},\tilde{\bp})}\P_i\neq 0)\\
&(\rho(\bx,\bp)=\rho(\tilde{\bx},\tilde{\bp})=\rho_i).
\end{split}
\]
Assuming that this holds, we can write
\[
\rho\P_i=\sum_{(\bx,\bp)}\rho(\bx,\bp)\P_{(\bx,\bp)}\P_i=\rho_i \sum_{(\bx,\bp)}\P_{(\bx,\bp)}\P_i=\rho_i\P_i,
\]
where we have used $\sum_{(\bx,\bp)}\P_{(\bx,\bp)}=1$. Summing the above equation over $i$, and using $\sum_{i}\P_{i}=1$, we obtain
\[
\rho=\sum_i\rho_i\P_i,
\]
i.e., $\rho$ can be decomposed using coarse-graining $\C=\{\P_i\}$.
Defining sets $I^{(\lambda)}=\{i|\rho_i=\lambda\}$,
we can rewrite this equation as
\[
\rho=\sum_\lambda\lambda\sum_{i\in I^{(\lambda)}}\P_i,
\]
and since decomposition Eq.~\eqref{eq:uniquelambdadecomposition} is unique, it must be that 
\[\label{eq:lastP}
\P_\lambda=\sum_{i\in I^{(\lambda)}}\P_i,
\]
which by definition means $\C_\rho\hookrightarrow \C$. For multiple corse-graining this then means that $\C_\rho\hookrightarrow \C_{1,\dots,n}$.

Conversely, we assume that Eq.~\eqref{eq:lastP} holds. Points $(\bx,\bp),(\tilde{\bx},\tilde{\bp})$ such that $\P_{(\bx,\bp)}\P_i\neq 0,\P_{(\tilde{\bx},\tilde{\bp})}\P_i\neq 0$ belong into the same macrostate $\P_i$, and therefore by Eq.~\eqref{eq:lastP} they must have the same associated eigenvalue,
\[
\rho(\bx,\bp)=\rho_i=\rho(\tilde{\bx},\tilde{\bp}),
\]
which means that Eq.~\eqref{eq:conditionrewritten} holds, thus inequality in \eqref{seriesofinequalitiesforgibbs} becomes equality.
\end{proof}

\begin{proof} (Theorem~\ref{thm:non-increase}) (Observational entropy is non-increasing with each added coarse-graining.)
Since joint coarse-graining $\C_{1,\dots,n}$ is finer than joint coarse-graining $\C_{1,\dots,n-1}$, from Theorem~\ref{thm:monotonic} we have
\[
S_{O(\C_{1,\dots,n})}(\rho)\leq S_{O(\C_{1,\dots,n-1})}(\rho).
\]
Statement of the theorem then follows from Eq.~\eqref{eq:joint_coarse}.
\end{proof}

\begin{proof}(Theorem~\ref{thm:stationary_states})
Using $\hat{P}_E \hat{P}_{\bn \tilde E}=\delta_{E,\tilde E}\hat{P}_{\bn \tilde E}$ we have
\[
\begin{split}
p_{\bn \tilde E}(\rho)&=\mu\int_{(\bx,\bp)} \sum_E f(E)\hat{P}_E(\bx,\bp) \hat{P}_{\bn \tilde E} (\bx,\bp) d\bx d\bp\\
&=\mu\int_{(\bx,\bp)} \bigg(\sum_E f(E)\delta_{E,\tilde E}\bigg)\hat{P}_{\bn \tilde E} (\bx,\bp) d\bx d\bp\\
&=f(\tilde E)\int_{(\bx,\bp)} \mu\hat{P}_{\bn \tilde E} (\bx,\bp) d\bx d\bp\\
&=f(\tilde E)V_{\bn \tilde E},
\end{split}
\]
from which
\[
\begin{split}
S_{xE}(t)&=-\sum_{\bn, E}f(E)V_{\bn E}\ln\frac{f(E)V_{\bn E}}{V_{\bn E}}\\
&=-\sum_{E}\Big(\sum_{\bn}V_{\bn E}\Big)f(E)\ln f(E)\\
&=-\sum_{E}V_{E}f(E)\ln f(E).
\end{split}
\]
\end{proof}

\section{Details for dynamics of $S_{xE}$ for a microcanonical ensemble}~\label{app:evolution}

Here we provide details for the calculation done in subsection~\ref{sec:starting_in_microcanonical}, showing that $S_{xE}=S_{O(\C_x,\C_{H_2})}$ for a microcanonical state of particles confined in the left part of the box is equal to the microcanonical entropy of this left part.

The initial state is a microcanonical state of $N$ particles contained in the left part of the box. It can be written in two ways as
\[
\rho_0\equiv\rho^{(\mathrm{micro};H_1)}_{E,\Delta E}=\frac{\mu\,\hat{P}_{E,\Delta E}^{(H_1)}}{V_{E,\Delta E}^{(H_1)}}=\frac{\mu\,\hat{P}_{(N,0)}\hat{P}_{E,\Delta E}^{(H_2)}}{V_{E,\Delta E}^{(H_1)}}.
\]
$\hat{P}_{E,\Delta E}^{(H_1)}$ and $\hat{P}_{E,\Delta E}^{(H_2)}$ denote projectors onto energy macrostates given by the first and the second Hamiltonian, and $\hat{P}_{(N,0)}$ is the projector onto a local particle number macrostate, corresponding to a statement that $N$ particles are in the left side of the box, and zero on the right.

We take $S_{xE}=S_{O(\C_x,\C_{H_2})}$ with local particle number coarse-graining that halves the box ($\bn=(n_1,n_2)$, $L_1=0$, $\Delta x=\frac{L}{2}$), and that has energy macrostates given by the second Hamiltonian, $H_2$. Using
\begin{align}
\hat{P}_{\bn}\hat{P}_{(N,0)}&=\delta_{n_1,N}\hat{P}_{(N,0)},\\
\hat{P}_{\tilde E}^{(H_2)}\hat{P}_{E,\Delta E}^{(H_2)}&=\begin{cases}
      \hat{P}_{(N,0)}\hat{P}_{\tilde E}^{(H_2)}, & \tilde{E}\in [E,E+\Delta E) \\
      0, & \mathrm{otherwise}
   \end{cases}
\end{align}
where $\delta_{n_1,N}$ denotes Kronecker delta, we have
\[
p_{\bn \tilde{E}}(\rho_0)=\begin{cases}
      \frac{V_{\tilde E}^{(H_1)}}{V_{E,\Delta E}^{(H_1)}}, & n_1=N\ \wedge\ \tilde{E}\in [E,E+\Delta E) \\
      0, & \mathrm{otherwise}
   \end{cases}
\]
from which we obtain
\[\label{eq:initial_micro}
S_{xE}(\rho_0)=\ln V_{E,\Delta E}^{(H_1)}\equiv S_{\mathrm{micro}}^{(H_1)}(E).
\]
In other words, $S_{xE}$ of the initial state is equal to the microcanonical entropy of the first half of the box.

\section{Dynamics of $S_{xE}$ for a single point in phase-space}\label{app:single_point}

Here we provide calculation on the long-time limit of $S_{xE}$ in the case of an initial single point in phase-space. We note that the derivation of $S_{xE}$ of an initial state follows exactly the same pattern as in Appendix~\ref{app:evolution}, with the only difference that we have $dE$ instead of $\Delta E$.

We now focus on calculating microcanonical entropy of the largest macrostate of phase-space jointly coarse grained by $\C_X$ and $\C_{H_2}$ (we will further denote $H\equiv H_2$).

Consider $N$ identical classical particles in $d$ dimensions evolving through a Hamiltonian $H$.
Their positions and momenta are described by a point in phase-space $\Gamma \equiv (\bx_1,\dots,\bx_N,\bp_1,\dots,\bp_N)$.
They are confined to a box of spatial volume ${\cal V}=L^d$, where $L$ is a linear dimension.
We subdivide the box into smaller ones of linear dimensions $\Delta x$, so that there are
$m\equiv \big(\frac{L}{\Delta x}\big)^d$ boxes. An arbitrary $\Gamma$ will be in some coarse-grained region $P_{\bn E}$,
and this will have $n_1$ particles in
the first box, $n_2$ in the second, and in general $n_i$ in the $i$'th box.
We would like to know how many distinct ways, $\CN$, there are of arranging the $N$ particles having
precisely these $\bn=(n_1,\dots,n_m)$. The answer is the multinomial distribution
\[
\CN(n_1,\dots,n_m) = \frac{N!}{\prod_{i=1}^m n_i!},
\]
where $\sum_{i=1}^m n_i = N$.

$\CN$ is maximimized when the $n_i$'s are uniformly distributed, that is, for large $\frac{N}{m}$, $n_i = \frac{N}{m}$.

The spatial volume of a coarse grained region is $\Delta x^{Nd} \CN(\{n_1,\dots,n_m\})$. If we fine
grain in energy, we can write the (phase-space) volume of the $i$th region that $\Gamma$ may be in:
\[\label{eq:Vi}
V_{\bn E} = \int dp^{Nd} \int_{x\in P_{\bn E}} dx^{Nd} \delta\big(E-H(\{\bx_i\}_i,\{\bp_i\}_i)\big),
\]
where $\hat{P}_E(\{\bx_i\}_i,\{\bp_i\}_i)= \delta\big(E-H(\{\bx_i\}_i,\{\bp_i\}_i)$ is the projector onto energy shell.
$P_{\bn E}$ contains all permutations of particles. This means that if we consider a
single $\Gamma$ in this region,  $P_{\bn E}$ will also contain other $\Gamma$'s with all of particle permutations consistent the same particle numbers, $\bn=(n_1,\dots,n_m)$, and the same energy $E$.

\subsection{Case of dilute gas}
Consider the case where the interaction between the particles is small. We will also take the particles to be indistinguishable, meaning all phase-space volumes are divided by $N!$.
In the limit of small interactions, the Hamiltonian only depends on the $\bp$s and we can integrate over $x$, and write
\[
V_{\bn E} =\frac{1}{N!} \frac{N!}{\prod_{i=1}^m n_i!}\Delta x^{Nd} \int dp^{Nd}  \delta\big(E-H(\{\bp_i\}_i)\big).
\]
Defining the number of microstates of $N$ indistinguishable particles in energy shell $E$ occupying spatial volume ${\cal V}$ as
\[
\Omega(E,{\cal V},N)={\cal V}^{N} \int dp^{Nd} \frac{1}{N!} \delta\big(E-H(\{\bp_i\}_i)\big),
\]
we can write
\[\label{eq:VnE}
V_{\bn E} = \frac{N!}{\prod_{i=1}^m n_i!}\Omega(E,\Delta x^d,N).
\]
$\Omega(E,{\cal V},N)$ defines the microcanonical entropy,\footnote{We remind that the microcanonical entropy is defined as by the right hand side of Eq.~\eqref{eq:microcanonical_entropy}, in which notation $V_E\equiv \Omega(E,{\cal V},N)\equiv \rho(E)\Delta E$. Here we use $\Omega(E,{\cal V},N)$ instead of the other two to spell out its dependence on the three variables.} $S_{\mathrm{micro}}(E,{\cal V},N)=\ln[\Omega(E,{\cal V},N)]$ (taking $k_B=1$). For small interactions considered here, we can approximate (Eq.~(2.40) in~\cite{schroeder1999introduction})
\[\label{eq:omega}
\Omega(E,{\cal V},N)=\frac{1}{N!}\frac{{\cal V}^N(2\pi M E)^{\frac{Nd}{2}}}{(\frac{Nd}{2})!},
\]
where $M$ is the mass of a gas particle.\footnote{This expression leads to the famous Sackur-Tetrode equation~\cite{ma1985statistical},
$S_{\mathrm{micro}}(E,{\cal V},N) = N\bigg(\ln\bigg(\frac{{\cal V}}{N}\bigg(\frac{4\pi M E}{N d}\bigg)^{\frac{d}{2}}\bigg) + 1 + \frac{d}{2}\bigg)
$, but we will not use it here directly.}

$\CN$ is maximized with $n_i = \frac{N}{m}$. In this case, using Stirling's approximation,
\[\label{eq:lnN}
\begin{split}
\ln(\CN_{\max}) &=\ln \frac{N!}{\Big(\frac{N}{m}!\Big)^m}\\
&\approx N\ln(m)  + \frac{1}{2} \ln (2\pi N)  -\frac{m}{2}\ln\Big(2\pi \frac{N}{m}\Big).
\end{split}
\]

Combining Eqs.~\eqref{eq:VnE},~\eqref{eq:omega}, and~\eqref{eq:lnN}, and realizing that $\Delta x^d=\frac{{\cal V}}{m}$ (where ${\cal V}$ denotes the full spatial volume), we obtain $S_{xE}$ for the point in phase-space $\Gamma$ which wandered into the largest macrostate,
\[
\begin{split}
S_{xE}&=\ln(V_{\bn E}^{(\max)})=\ln(\CN_{\max})+\ln\Omega(E,\Delta x^d,N)\\
&=\ln(\CN_{\max})+\ln( m^{-N}\Omega(E,{\cal V},N))\\
&\approx N\ln(m) + \frac{1}{2} \ln (2\pi N)  -\frac{m}{2}\ln\Big(2\pi \frac{N}{m}\Big)\\
&-N\ln(m)+\ln \Omega(E,{\cal V},N)\\
&=S_{\mathrm{micro}}(E,{\cal V},N)+ \frac{1}{2} \ln (2\pi N)  -\frac{m}{2}\ln\Big(2\pi \frac{N}{m}\Big).
\end{split}
\]

If the point in phase-space did not wander into the largest macrostate, but, let's say, the second largest (given by $n_1=\frac{N+1}{m}-1, n_i=\frac{N+1}{m}$ for $i>1$),  there would be minor modifications to the above formula, which would be negligible compared to the first term - the microcanonical entropy. In the end, it does not matter into which of the large macrostate the particle wanders -- the leading term will be always the microcanonical entropy.

To understand to correction term in more detail, we can look at the entropy per particle.  Then the relevant quantity to
consider here is $\ln(\CN_{max})/N$.
\[
\frac{S_{xE}}{N} =  \frac{S_{\mathrm{micro}}(E,{\cal V},N)}{N}  + \frac{1}{2N} \ln (2\pi N)  -\frac{m}{2N}\ln\Big(2\pi \frac{N}{m}\Big).
\]
We take a limiting process where we fix the particle density, and send $N$ to infinity, while keeping the particle density
constant. For large $N$ we can ignore the second term on the right hand side. $N_m \equiv \frac{N}{m}$ is the average number of
particles per box. In the limit where we fix the density average number of particles and increase the box length so
that $N_m$ becomes large, the last term on the right hand side $-(1/2N_m)\ln(2\pi N_m)$ becomes
small, and vanishes as $N_m\rightarrow \infty$. This correction term therefore represents a finite size effect.

\subsection{General classical systems}

The case of a dilute gas can be extended to a general homogeneous classical system
that is extensive. The system with
maximum entropy will be the one where the number in each box is the same. This follows from the fact
that all boxes must have the same chemical potential in equilibrium. The system is most likely to
find itself in this coarse-grained region. The logarithm of Eq.~\eqref{eq:Vi} represents the entropy of a system that has
been partitioned. As compared to the entropy of the entire system, each particle is confined to
a box of width $\Delta x$. It is as if barriers had been added to prevent the exchange of particles between
regions. However energy can still be exchanged, therefore the temperature of each sub-system is the
same. If the system has a density $\rho$ and temperature $T$, corresponding to its total energy,
then the entropy per particles $s(E,T)$ will not depend on system size in the thermodynamic limit.
There are correction to this due to finite size effects and are particularly pronounced at critical
points~\cite{privman1984universal}. They have the general behavior as above, becoming negligible
in the limit of large box size. The actual error will depend on the universality class of the system
being studied.

\section{Properties of $S_F$}\label{app:propertiessf}

First we rephrase the coarse-graining $\C_{X\bE}=\{P_{\bn\bE}\}_{\bn,\bE}$, Eq.~\eqref{eq:localnlocalE}, in mathematically precise terms. For simplicity we set $L_1=0$ and $L_2=L$.

The local particle number-local energy macrostates are then
\[\label{eq:localnlocalEmath}
\begin{split}
&P_{\bn\bE}\equiv \{(\bx,\bp)|n_1 \text{ particles with position } x \in [0,\Delta x),\\
&\ \dots, n_m \text{ particles with position } x \in [(m\!-\!1)\Delta x,L),\\
& \text{energy of the bin } [0,\Delta x) \text{ is between } [E_1,E_1+dE),\\
&\dots,\text{energy of the bin } [(m\!-\!1)\Delta x,L)\text{ is between }\\ & [E_m,E_m+dE)\}.
\end{split}
\]

In order to define this in mathematical terms, we need to define the notion of \emph{local Hamiltonian}. While the general theory that applies also to a phase-space density will be done elsewhere~\cite{safranek2020classicalee}, here we can still treat the case of a single point of phase-space. At every time $t$, there there is some number of particles in each of the $m$ bins. Let's say we have $n_1$ particle between $[0,\Delta x)$. For this particular time, the local Hamiltonian $H_1$ will be a function of phase-space points of $n_1$ particles,\footnote{For a different time, it can be a function of a different number of particles -- this hints on the construction for the general case which requires the construction of a Fock phase-space~\cite{safranek2020classicalee}.}
which maps a point in phase-space to the energy\[
H_1:(x_{1},\dots,x_{n_1},p_{1},\dots,p_{n_1})\rightarrow E_1, 
\]
where $x_i\in[0,\Delta x)$,  $i=1,\dots,n_1$ (note that the lower index here is just a label denoting variables -- we are not saying that exactly the first $n_1$ particles are in the first bin; in case of indistinguishable particles, this label does not even matter). For instance, in the case of non-interacting particles we have $H_1=\sum_{i=1}^{n_1}\frac{p_i^2}{2M}$. If there is some interaction, the interaction between particles within the bin is accounted for in the local Hamiltonian, while the interaction between different bins is ignored.

We denote $S_N$ the set of permutations on the set $\{1,2,\dots,N\}$. With these tools, we can write
\[\label{eq:localnlocalEmath}
\begin{split}
&P_{\bn\bE}\equiv \{(\bx,\bp)|(\exists \pi\in S_N)\ \text{such that}\ \\
&(\forall x_{\pi(i)}|i=1,\dots,n_1)(x_{\pi(i)} \in [0,\Delta x)),\dots,\\
&(\forall x_{\pi(i)}|i=N-n_m+1,\dots,N)(x_{\pi(i)} \in [(m\!-\!1)\Delta x,L)),\\
&E_1\leq H_1(x_{\pi(i)},p_{\pi(i)})<E_1+dE,\dots,\\
&E_m\leq H_m(x_{\pi(i)},p_{\pi(i)})<E_m+dE\}.
\end{split}
\]
Clearly, independently of the permutation, the phase-space volume will be always the same for a specific permutation $\pi$. The question, then, is how many permutations lead to a specific distribution of particles $\bn$ between the bins. The answer is again
\[
\CN(n_1,\dots,n_m) = \frac{N!}{\prod_{i=1}^m n_i!},
\]
which is the total number of permutations, divided by the possible number of permutations within the first bin, within the second bin, and so on. We can therefore write the volume of macrostate $P_{\bn\bE}$ as
\[
\begin{split}
&V_{\bn\bE}=\CN(n_1,\dots,n_m)\frac{1}{N!}\Pi_{i=1}^{n_1}\int_{x_i\in[0,\Delta x)}dx_i\int_{p_i}dp_i\times\\
&\cdots\times\Pi_{i=N-n_m+1}^{N}\int_{x_i\in[(m-1)\Delta  x,L)}dx_i\int_{p_i}dp_i\times\\
&\delta(E_1-H_1(x_{1},\dots,x_{n_1},p_{1},\dots,p_{n_1}))\times\cdots\\
&\times\delta(E_m-H_m(x_{N-n_m+1},\dots,x_{N},p_{N-n_m+1},\dots,p_{N}))\\
&=\Pi_{k=1}^m\Omega(E_k,\Delta x, n_k).
\end{split}
\]
Assuming generalization to spatial dimension $d$, we similarly derive $V_{\bn\bE}=\Pi_{k=1}^m\Omega(E_k,\Delta x^d, n_k)$.
This means that for a single point in phase-space $\gamma\in P_{\bn\bE}$,
\[
S_F=\ln V_{\bn\bE}=\sum_{k=1}^m S_{\mathrm{micro}}(E_k,\Delta x^d,n_k),
\]
and therefore $S_F$ is additive.

Now, let us study the long-time limit. After a long time, assuming identical particles of a dilute and weakly interacting gas, the largest macrostate is given by $E_k=\frac{E}{m}$, $n_k=\frac{N}{m}$, (where $m=\frac{\cal V}{\Delta x^d}$), giving
\[
\begin{split}
S_F&=\ln(V_{\bn \be}^{(\max)})=m S_{\mathrm{micro}}(\tfrac{E}{m},\Delta x^d,\tfrac{N}{m})\\
&=m\ln\frac{1}{\tfrac{N}{m}!}\frac{(\tfrac{\cal V}{m})^{\frac{N}{m}}(2\pi M \tfrac{E}{m})^{\frac{Nd}{2m}}}{(\frac{Nd}{2m})!}\\
&=S_{\mathrm{micro}}(E,{\cal V},N)+\ln\bigg(\frac{m^{-N-\frac{Nd}{2}}N!\frac{Nd}{2}!}{(\frac{N}{m}!)^m(\frac{Nd}{2m}!)^m}\bigg)\\
&=S_{\mathrm{micro}}(E,{\cal V},N)+ \ln (2\pi N)  -m\ln\Big(2\pi \frac{N}{m}\Big)\\
&-\frac{m-1}{2}\ln\frac{d}{2}.
\end{split}
\]
Here we have used Eq.~\eqref{eq:omega} and  Stirling's approximation for the factorials. The correction term $\frac{m-1}{2}\ln\frac{d}{2}$ represents a finite-size effect, which result in an order-1 correction in the thermodynamic limit.

\begin{figure*}[!t]
\begin{center}
\includegraphics[width=0.49\hsize]{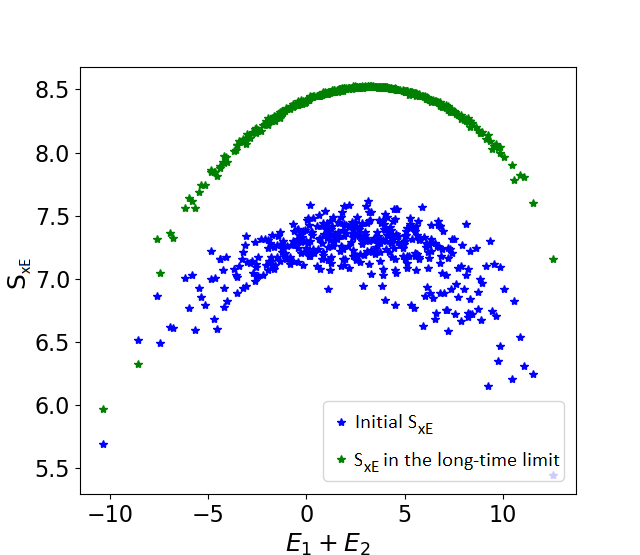}
\includegraphics[width=0.49\hsize]{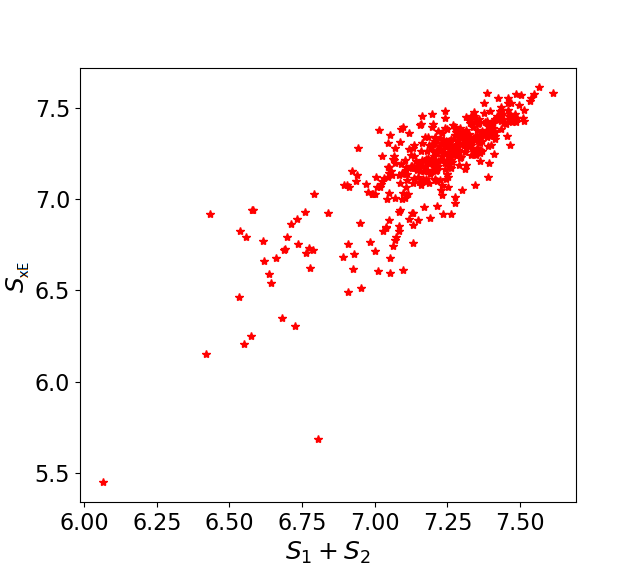}
\caption
{
Simulation showing additivity of quantum $S_{xE}$ with infinite resolution in energy ($\Delta E=0$). The system studied is of spinless fermions on a 1-dimensional lattice evolved by a non-integrable Hamiltonian used in~\cite{safranek2019letter,safranek2019long}. The parameters are a lattice of $L=16$ sites with hard-wall boundary conditions, with coarse-graining size $\Delta x=4$ sites and the total number of particles $N=6$. The state is initialized as a product state of random pure thermal states~\cite{safranek2019letter,safranek2019long}), $\ket{\psi}=\ket{\psi_{E_1}}\otimes\ket{\psi_{E_2}}$, where $\ket{\psi_{E_1}}$ is a state of $3$ particles occupying the $8$ sites on the left of the lattice with mean energy $E_1$, and $\ket{\psi_{E_2}}$ is a state of $3$ particles occupying the $8$ sites on the right with mean energy $E_2$. The left figure represents comparison of the $S_{xE}$ of the initial state, $S_{xE}(\ket{\psi})$, with the $S_{xE}$ in the long-time limit, $S_{xE}(e^{-i\hat{H}t}\ket{\psi})$ (with a large $t$). Classically, these two values would be the same and the curves would overlap, because both would corresponds to $S_{xE}=\ln V_{(3,3)E}$ (where $E=E_1+E_2$). As we can see, they do not overlap in a quantum mechanical system when $\Delta E=0$. The right figure shows the additivity of $S_{xE}$ directly, by comparing $S_{xE}$ and the sum of thermodynamic entropies of the left $8$ sites ($S_1$) and of the right $8$ sites ($S_2$), for various product states $\ket{\psi}=\ket{\psi_{E_1}}\otimes\ket{\psi_{E_2}}$ such that $E=E_1+E_2$.
}
\label{Fig:additivitySxE}
\end{center}
\end{figure*}

\section{Additivity of quantum $S_{xE}$ and non-additivity of classical}\label{app:additivityofsxe}

As explained in Section~\eqref{sec:interpretationofSxE}, the classical $S_{xE}$ is not additive. This is because the classical global energy macrostate is defined as 
\[
P_E\equiv \{(\bx,\bp)|E\leq H(\bx,\bp)<E+\Delta E\}
\]
This means that any microstate $(\bx,\bp)$ that has energy $E_1$ in the left region and energy $E_2$ in the right region, $E_1+E_2=E$, will fall into the same macrostate $P_E$ with any other states given by $\tilde E_1+\tilde E_2=E$.

Quantum mechanically, for infinite resolution in energy ($\Delta E=0$), this is not true. This is because in a generic system, even states with very similar sum of energies $E\approx E_1+E_2\approx \tilde E_1+\tilde E_2$ are not exactly the same, so energy eigenstates of the global Hamiltonian $\ket{E_1+E_2}$ and $\ket{\tilde E_1+\tilde E_2}$ constitute their own macrostates, $\hat{P}_{E_1+E_2}=\pro{E_1+E_2}{E_1+E_2}$ and $\hat{P}_{\tilde E_1+\tilde E_2}=\pro{\tilde E_1+\tilde E_2}{\tilde E_1+\tilde E_2}$. (We ignored corrections coming from ignoring the interaction terms between the two regions that also contribute to the total Hamiltonian and therefore also to local energy eigenstates.)

If a finite resolution in energy is used ($\Delta E>0$), then the situation for quantum $S_{xE}$ is very similar to the classical case, because then the quantum macrostates are constituted by
\[
\P_{E,\Delta E}\approx\sum_{E\leq E_1+E_2\leq E+\Delta E}\P_{E_1+E_2}.
\]
Put simply, classical energy macrostates correspond to quantum macrostate with $\Delta E>0$, but for $\Delta E=0$, the quantum macrostates are much smaller -- actually consisting of a single microstate. This means that for $\Delta E=0$ the global energy macrostates are actually small enough to distinguish between different local energies, which is why the quantum $S_{xE}$ is additive.

We have confirmed this numerically in some particular cases, as exhibited in Fig.~\ref{Fig:additivitySxE}.

Making the analytical argument for additivity precise would require similar arguments to those employed in appendices of Ref.~\cite{safranek2019long}. That is, identifying the interaction part of the Hamiltonian with a random matrix, and then studying the overlaps between energy eigenstates the full Hamiltonian of the system and those of the Hamiltonian system without the interaction, which gives the probability $p_{E}=\abs{\braket{E_1,E_2}{E}}^2$ that goes into $S_{xE}$. We leave this for future work.

\bibliographystyle{apsrev4-1}
\bibliography{class_OEBib}

\end{document}